\documentclass[letterpaper]{jpconf}
\usepackage{graphicx}
\begin{document}
\title{Transverse Momentum Correlations in Relativistic Nuclear Collisions}

\author{Thomas A Trainor and Duncan J Prindle (STAR Collaboration)}

\address{CENPA 354290, University of Washington, Seattle, WA  98195 USA}

\ead{trainor@hausdorf.npl.washington.edu,prindle@npl.washington.edu}

\def\bea{\begin{eqnarray}}
\def\eea{\end{eqnarray}}

\begin{abstract}
From the correlation structure of transverse momentum $p_t$ in relativistic nuclear collisions we observe for the first time temperature/velocity structure resulting from low-$Q^2$ partons. Our novel analysis technique does not invoke an {\em a priori} jet hypothesis. $p_t$ autocorrelations derived from the scale dependence of $\langle p_t \rangle$ fluctuations reveal a complex parton dissipation process in RHIC heavy ion collisions. We also observe structure which may result from collective bulk-medium recoil in response to parton stopping.
\end{abstract}.

\section{Introduction}

Central Au-Au collisions at RHIC may generate a color-deconfined medium 
(quark-gluon plasma or QGP)~\cite{QCD}. Some  theoretical descriptions predict abundant low-$Q^2$ gluon production in the early stages of high-energy nuclear collisions, with rapid parton thermalization as the source of the colored medium~\cite{theor0,theor1,theor2}. Nonstatistical fluctuations of event-wise mean $p_t$ $\langle p_t \rangle$ \cite{Phenix,ptprl} may isolate fragments from low-$Q^2$ partons and determine the properties of the corresponding medium. A recent measurement of excess $\langle p_t \rangle$ fluctuations in Au-Au collisions at 130 GeV revealed a large excess of fluctuations compared to independent-particle $p_t$ production~\cite{ptprl}. 

In this paper we describe the event-wise structure of transverse momentum $p_t$ produced in relativistic nuclear collisions at RHIC. 
We discuss the role of low-$Q^2$ partons as Brownian probe particles in heavy ion collisions. We compare joint autocorrelations on $(\eta,\phi)$ to conventional leading-particle techniques for parton fragment analysis. We present experimental evidence from mean-$p_t$ fluctuations and corresponding $p_t$ autocorrelations for local temperature/velocity structure in A-A collisions which can be interpreted in terms of parton dissipation in the A-A medium and same-side recoil response of the bulk medium to parton stopping. Finally, we review the energy dependence of mean-$p_t$ fluctuations from SPS to RHIC and its implications.

\section{Low-$Q^2$ partons as Brownian probes}

\begin{figure}[h]
\begin{minipage}{38pc} \hfil
\includegraphics[width=24pc,height=8.5pc]{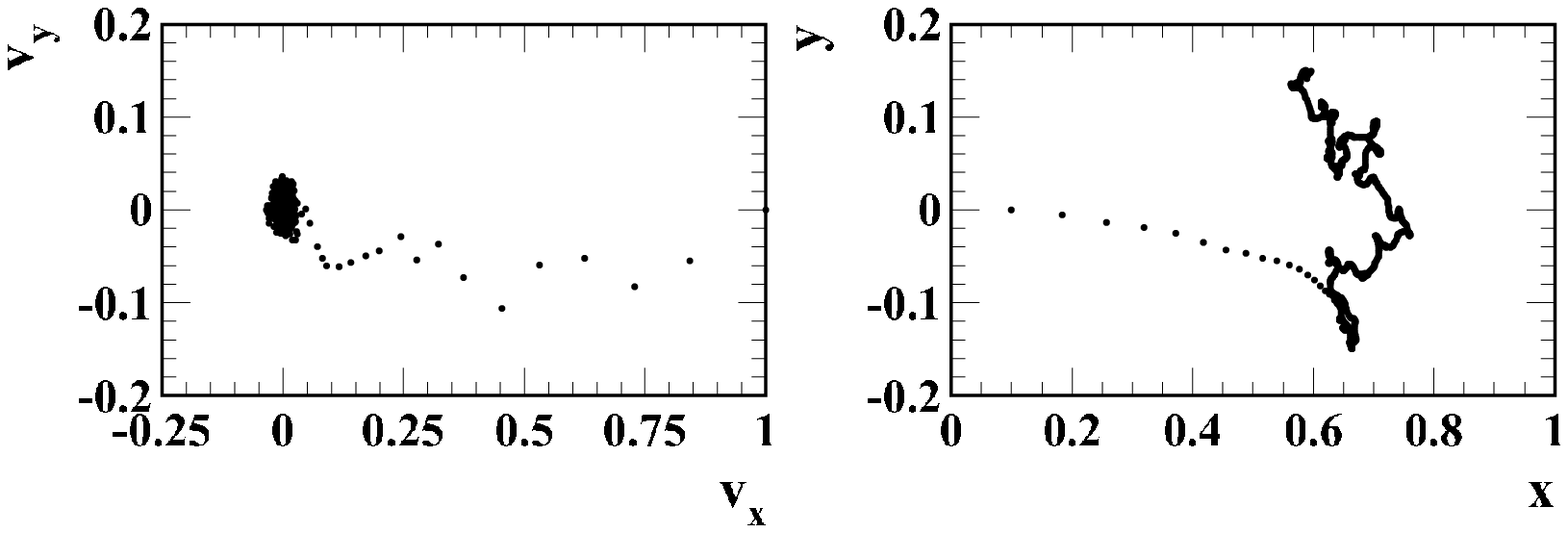} \hfil
\includegraphics[width=11pc,height=8.5pc]{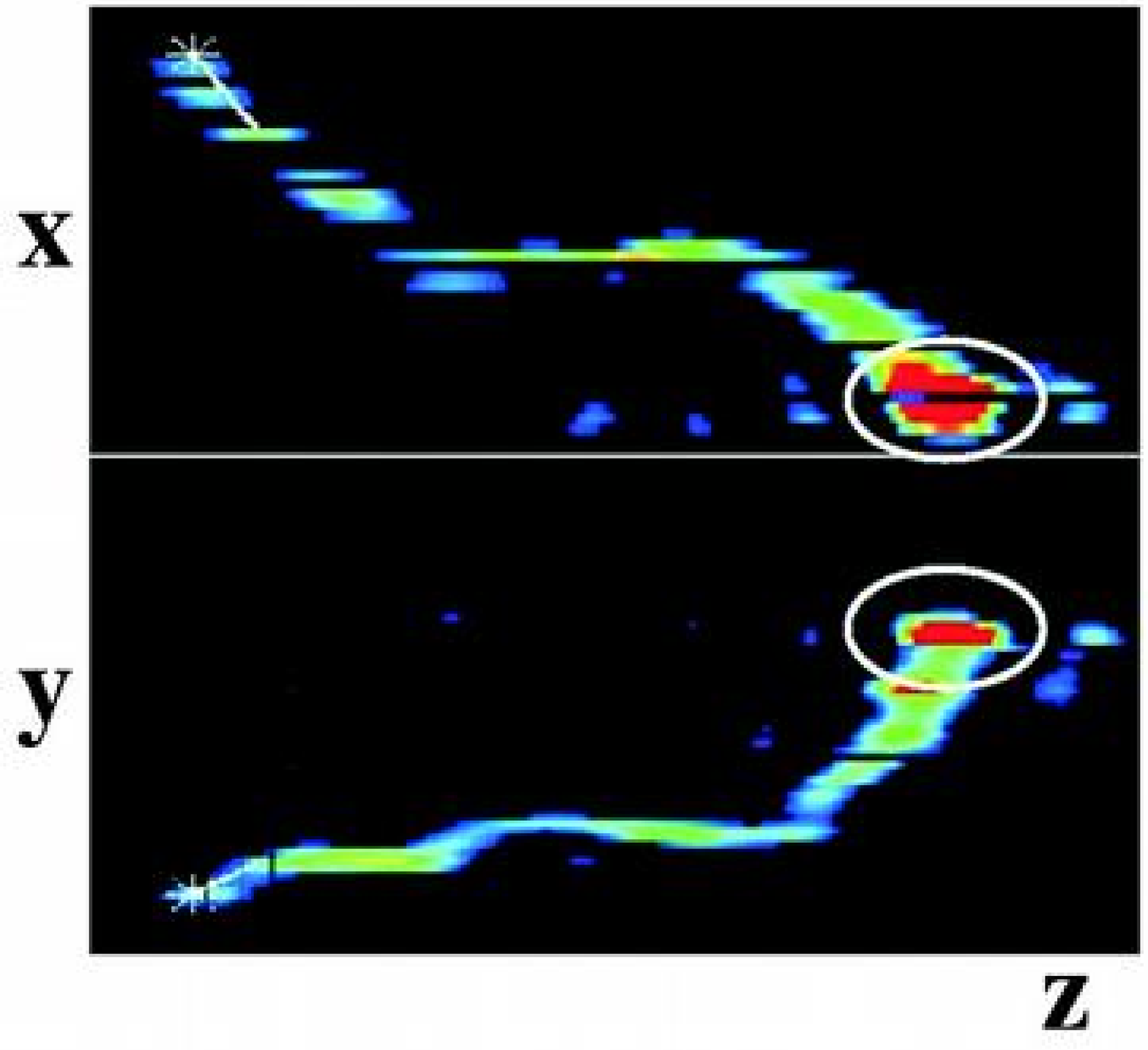} \hfil
\end{minipage}%
\caption{\label{langevin}  Left panels: Solution to the Langevin equation for a point mass with unit initial velocity in the positive $x$ direction illustrating thermalization. Right panel: two views of an electron track in a TPC, with charge ball from random walk at track endpoint (circled)~\cite{beta}.}
\end{figure}

In 1905 the microscopic structure of ordinary matter was addressed theoretically by Einstein, who introduced the concept of a (Brownian) probe particle large enough to be observed visually, yet small enough that its motion in response to the molecular dynamics of a fluid might also be observed~\cite{brown}. Those two constraints specified the one-micron probe particles used by Jean Perrin to confirm molecular motion in fluids~\cite{perrin,haw}. The Langevin equation $\dot{\vec{v}}(t) = -\frac{1}{\tau}\, \vec{v}(t) + \vec{a}_{stoch}(t) + \vec{a}_{mcs}(t)$ models the motion of a Brownian probe in a thermalized fluid medium of point masses qualitatively smaller than the probe particle~\cite{langevin,lemons}. The accelerations are gaussian-random with zero mean; $\vec{a}_{stoch}(t)$ is isotropic and $\vec{a}_{mcs}(t) \perp \vec{v}(t)$ (and $\propto v$). The first term models collective dissipation of probe motion (viscosity), the second models individual probe collisions with medium particles and the third simulates multiple Coulomb scattering of a fast probe particle. A solution of that equation for unit initial speed in the $x$ direction starting at the ($x,y$) origin is shown in the first two panels of Fig.~\ref{langevin}. Speed is dissipated with time, leading to equilibration with the medium: fluctuations of velocity about zero and random walk of the probe. An example of such motion is shown in the third panel: an electron track in a time projection chamber exhibits multiple Coulomb scattering along its trajectory, terminating in random walk represented by the circled ball of charge at the trajectory endpoint~\cite{beta}.

In 2005 we seek the microscopic structure and local dynamics of the QCD medium formed in RHIC heavy ion collisions. The point-mass concept of Einstein's Brownian probe must be extended to partonic probes, possibly with internal degrees of freedom and experiencing complex non-point interactions with medium degrees of freedom. This problem requires novel analysis techniques closely coupled to the Langevin equation and its associated numerical methods. The analog in heavy ion collisions to Einstein's Brownian probe is the low-$Q^2$ parton, visualized for the first time by methods presented in this paper. 
In contrast to Einstein's notion of a particle of exceptional size observed indefinitely in equilibrium with microscopic motions of a thermalized particulate medium, the QCD Brownian probe is identical to medium particles but possesses an exceptional initial velocity relative to the medium with which it interacts for a brief interval: do probe manifestations in the hadronic system reveal `microscopic' degrees of freedom of the medium, is the medium locally or globally equilibrated, what are its fluid properties?

\section{Joint autocorrelations {\em vs} conditional distributions}


Conventional study of QCD jets in elementary collisions is inherently model-dependent. Scattered partons with large transverse momentum are associated individually with concentrations of transverse momentum or energy localized on angle variables $(\eta,\phi)$. In heavy ion collisions, where such identification is impractical, jet studies are based on a high-$p_t$ `leading particle' which may estimate a parton momentum direction and some fraction of its magnitude. The leading-particle momentum is the basis for two-particle {\em conditional distributions} on transverse momentum and angles. Those distributions reveal medium modifications to parton production and fragmentation as changes in the single-particle $p_t$ spectrum ($R_{AA}$) and in the fragment-pair relative azimuth distribution (away-side jet disappearance), referred to collectively as {\em jet quenching}~\cite{raa-away}. The leading-particle approach is based on perturbative concepts of parton hard scattering as a point-like binary interaction and parton energy loss as gluon bremsstrahlung. We can then ask how the medium is modified by parton energy loss and what happens to {\em low}-$Q^2$ partons, in a $Q^2$ regime where the pQCD assumption of point-like interactions breaks down, where the parton may have an effective internal structure. In other words, how can we describe parton dissipation as a transport process, including bulk-medium degrees of freedom?

\begin{figure}[h]
\begin{minipage}{17pc}
\includegraphics[width=18pc,height=16pc]{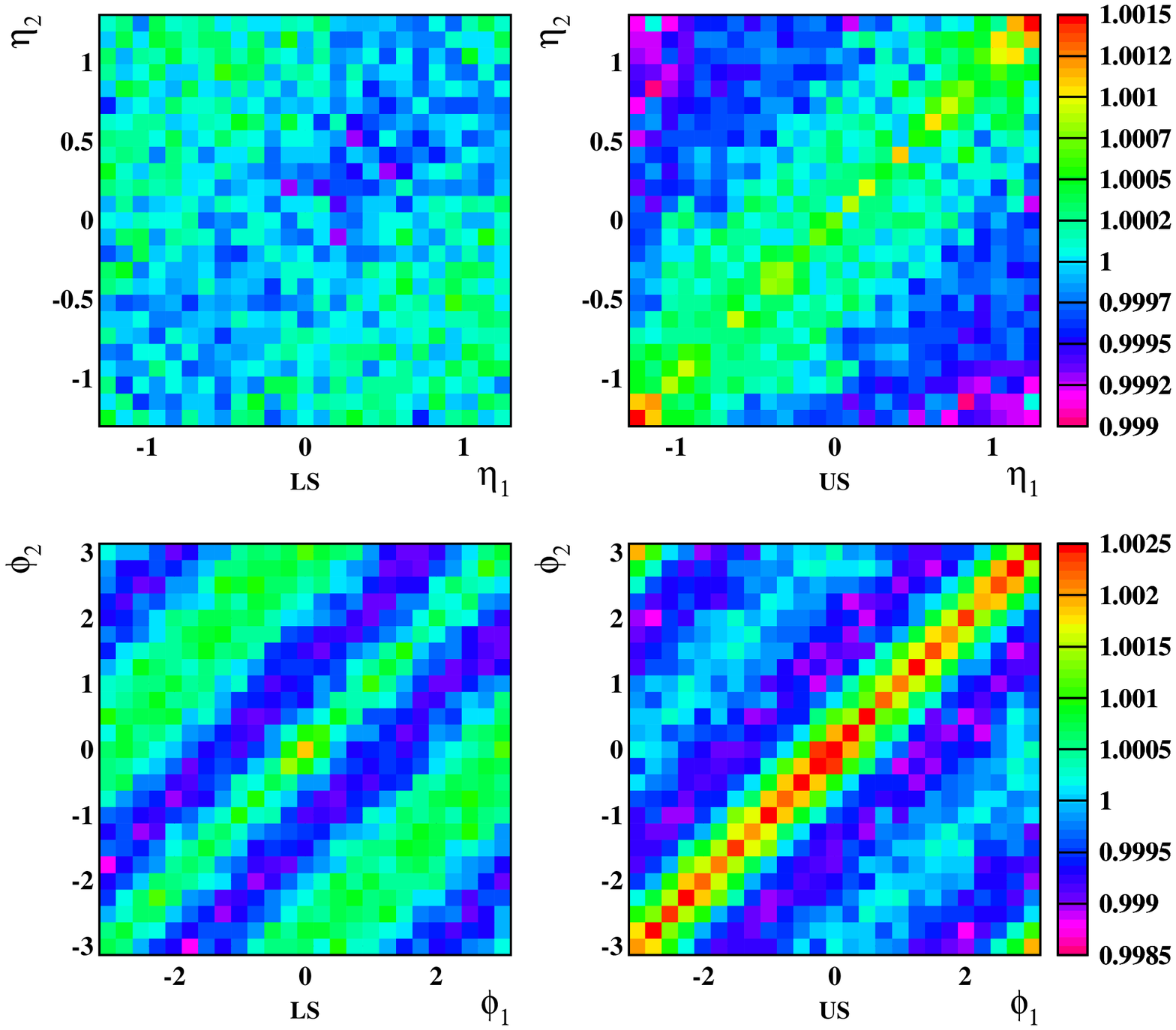}
\end{minipage}\hspace{0pc}%
\hfil
\begin{minipage}{9pc}
\includegraphics[width=9.5pc,height=8pc]{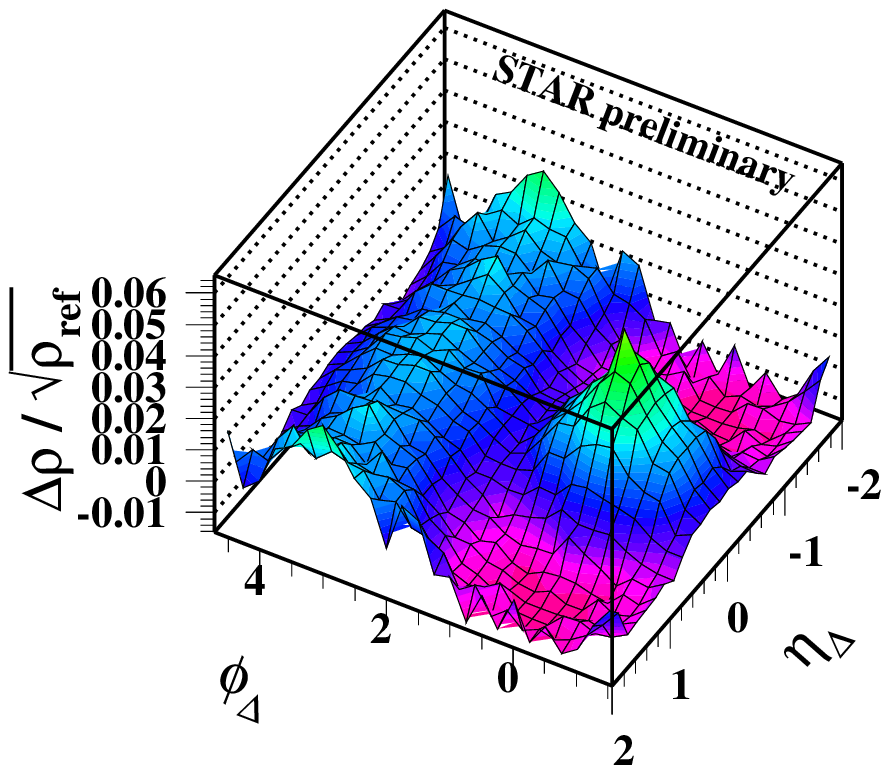} 
 \includegraphics[width=9.5pc,height=8pc]{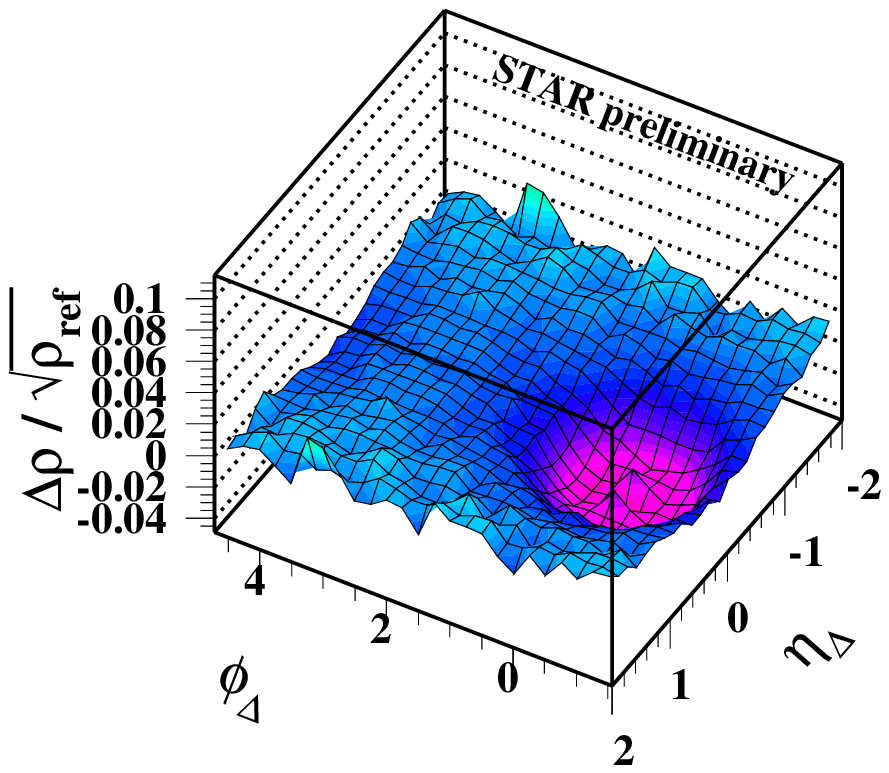} 
\end{minipage}\hspace{1pc}%
\begin{minipage}{8pc}
\includegraphics[width=9.pc,height=8pc]{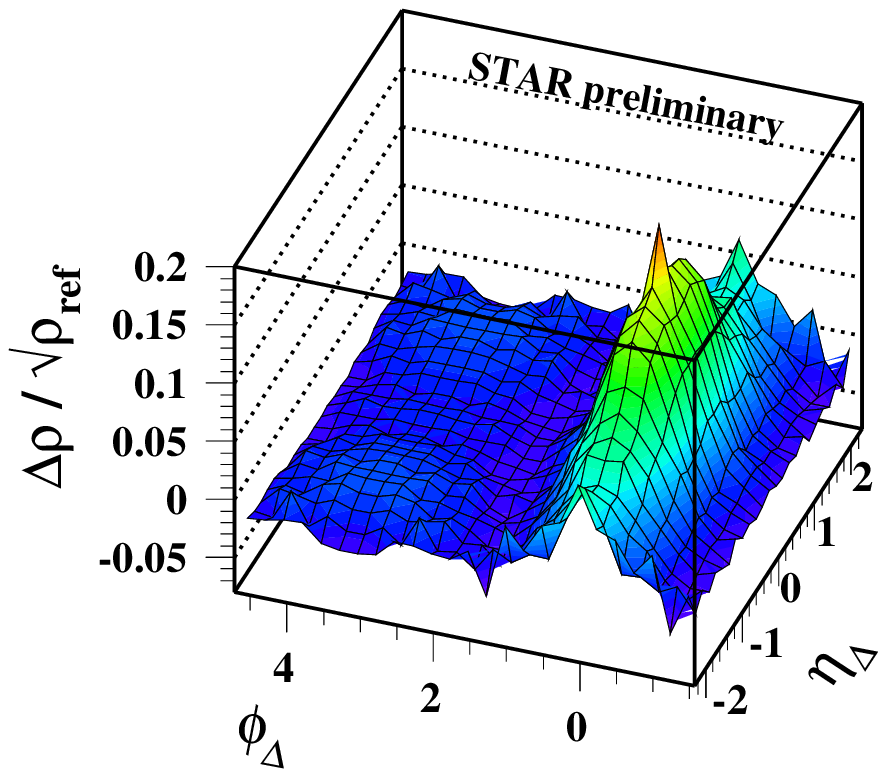} 
 \includegraphics[width=9pc,height=8pc]{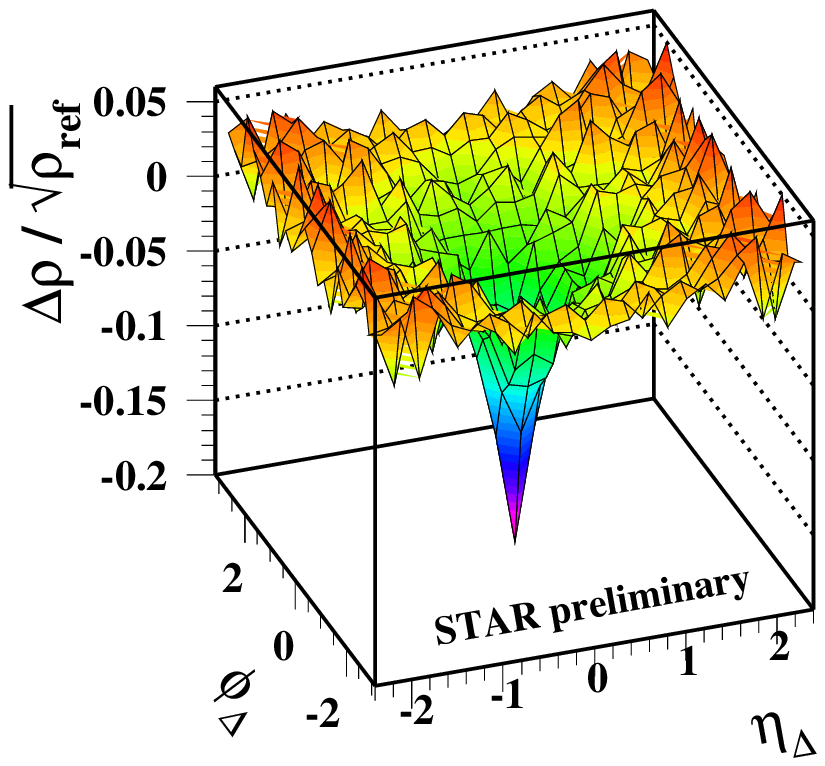} 
\end{minipage}\hspace{0pc}%
\caption{\label{auto}  Left panels: Two-particle distributions on $\eta$ and $\phi$ from central Au-Au collisions at 130 GeV illustrating stationarity along the sum axis. Right panels: Joint autocorrelations on $(\eta_\Delta,\phi_\Delta)$ from p-p collisions (left) and Au-Au collisions (right). 
}
\end{figure}

To access low-$Q^2$ partons we have developed an alternative analysis method for jet correlations employing {\em autocorrelation} distributions which do not require a leading- or trigger-particle concept. The autocorrelation principle is illustrated in Fig.~\ref{auto}. Projections of the two-particle momentum space of 130 GeV Au-Au collisions onto subspaces $(\eta_1,\eta_2)$ and $(\phi_1,\phi_2)$ (left panels) indicate that correlations on those spaces are approximately invariant on sum variables $\eta_\Sigma \equiv \eta_1 + \eta_2$ and $\phi_\Sigma \equiv \phi_1 + \phi_2$, in which case autocorrelations on difference variables $\eta_\Delta \equiv \eta_1 - \eta_2$ and $\phi_\Delta \equiv \phi_1 - \phi_2$ retain nearly all the information in the unprojected distribution~\cite{axialcd}. The autocorrelation concept was first introduced to solve the Langevin equation, to extract deterministic information from stochastic trajectories. In time-series analysis the autocorrelation of time series $f(t)$ is $A(\tau) = \frac{1}{T} \int_{-T/2}^{T/2} f(t)\, f(t+\tau)\, dt$, where difference variable $\tau \equiv t_1 - t_2$ is the {\em lag}. For a {\em stationary} distribution ($f(t)$ correlations statistically independent of absolute time) the autocorrelation represents a {\em projection by averaging} of all the information in $f(t)$. 

The same principle can be applied to ensemble-averaged two-particle momentum distributions which are approximately invariant on their sum variables~\cite{inverse}. Distributions on angle space $(\eta_1,\eta_2,\phi_1,\phi_2)$ can be reduced to {\em joint autocorrelations} on difference variables $(\eta_\Delta,\phi_\Delta)$. For example, joint autocorrelations 
in the right-most two panels of Fig.~\ref{auto} correspond to $(\eta_1,\eta_2)$ and $(\phi_1,\phi_2)$ distributions in the left-most four panels. Joint autocorrelations for relativistic nuclear collisions retain almost all correlation structure on a visualizable 2D space and provide access to parton fragment angular correlations with no leading-particle condition, sampling a {\em minimum-bias} parton distribution. Jet correlations are thus revealed with no {\em a priori} jet hypothesis, providing access to the low-$Q^2$ partons which serve as Brownian probes of the QCD medium.

\section{The p-p reference system}

\begin{figure}[h]
\begin{minipage}{18pc}
\begin{center}
 \includegraphics[width=8.5pc,height=8pc]{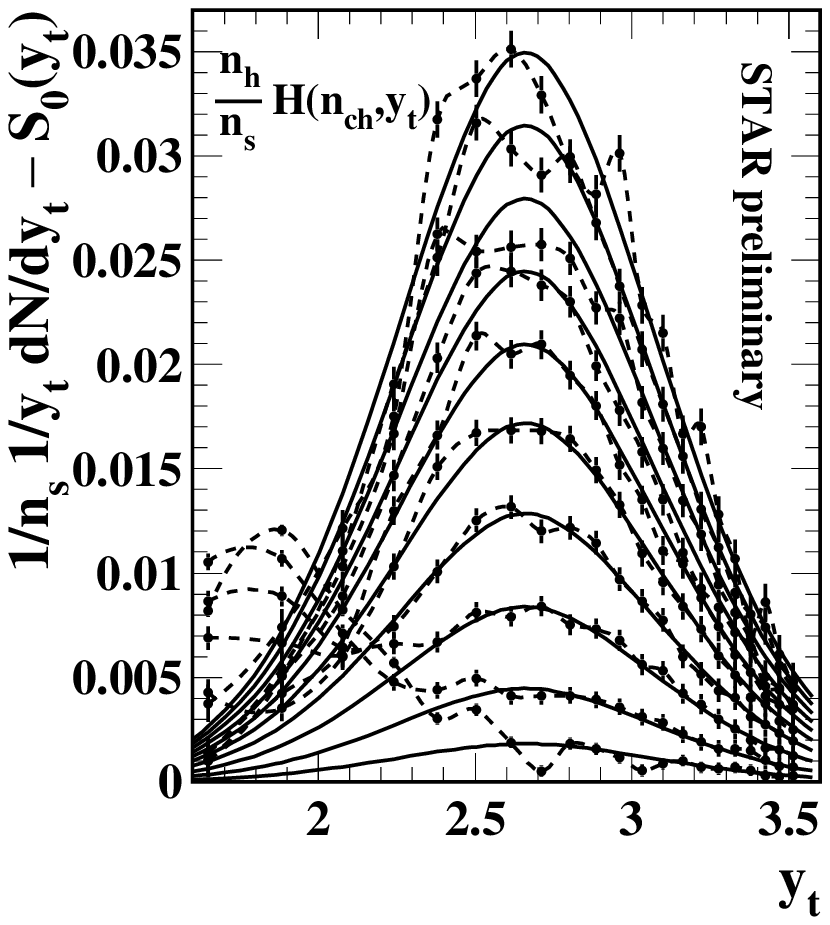} 
\includegraphics[width=8.5pc,height=8pc]{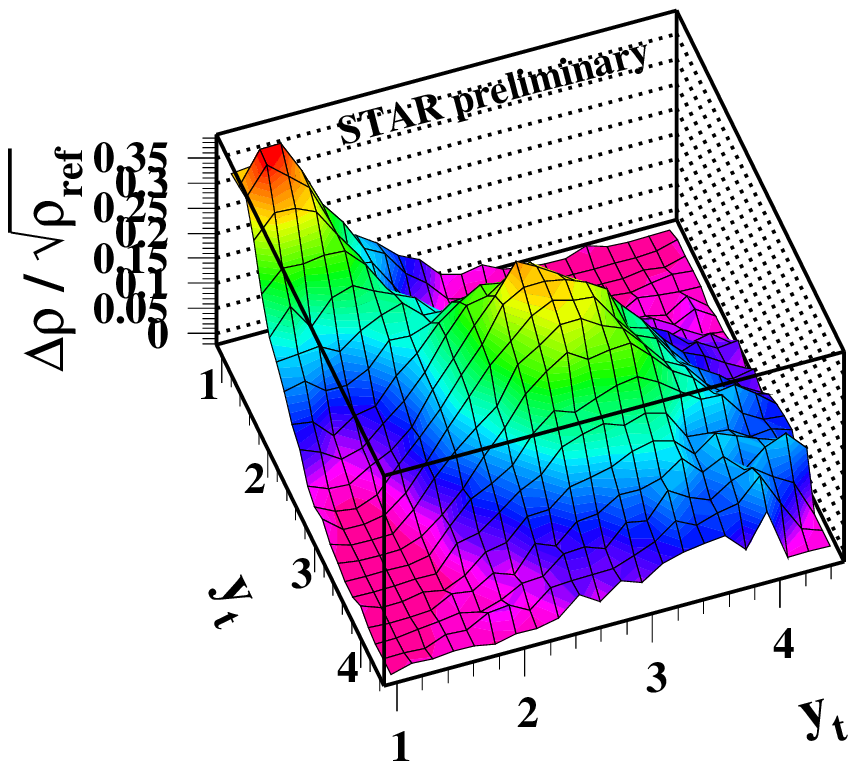}
\end{center} 
\end{minipage}
\hfil
\begin{minipage}{20pc}
\begin{center}
 \includegraphics[width=9.5pc]{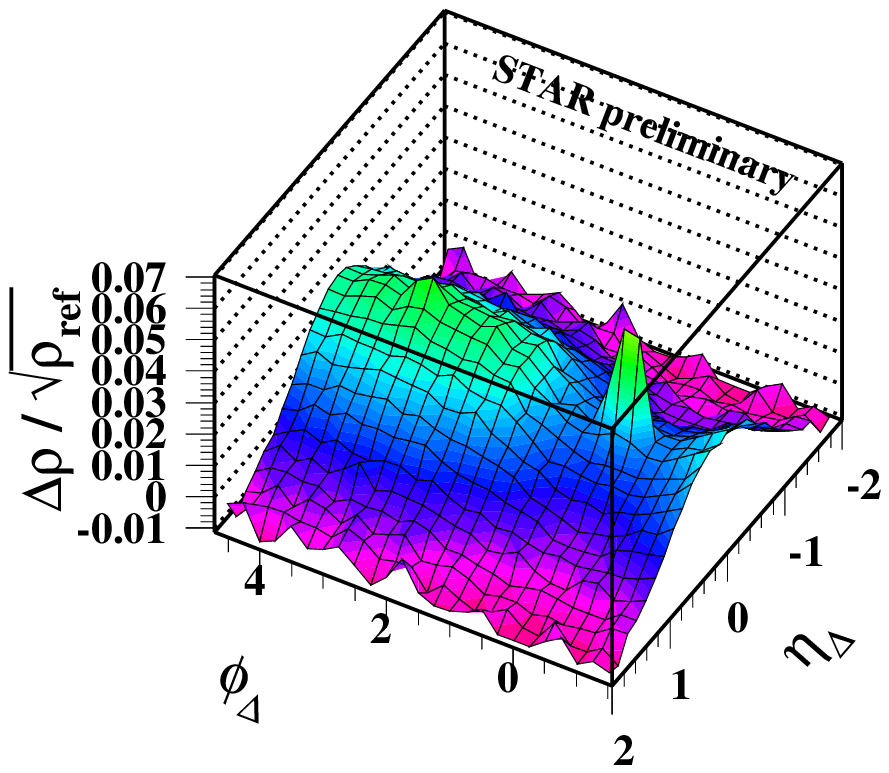} 
\includegraphics[width=9.5pc]{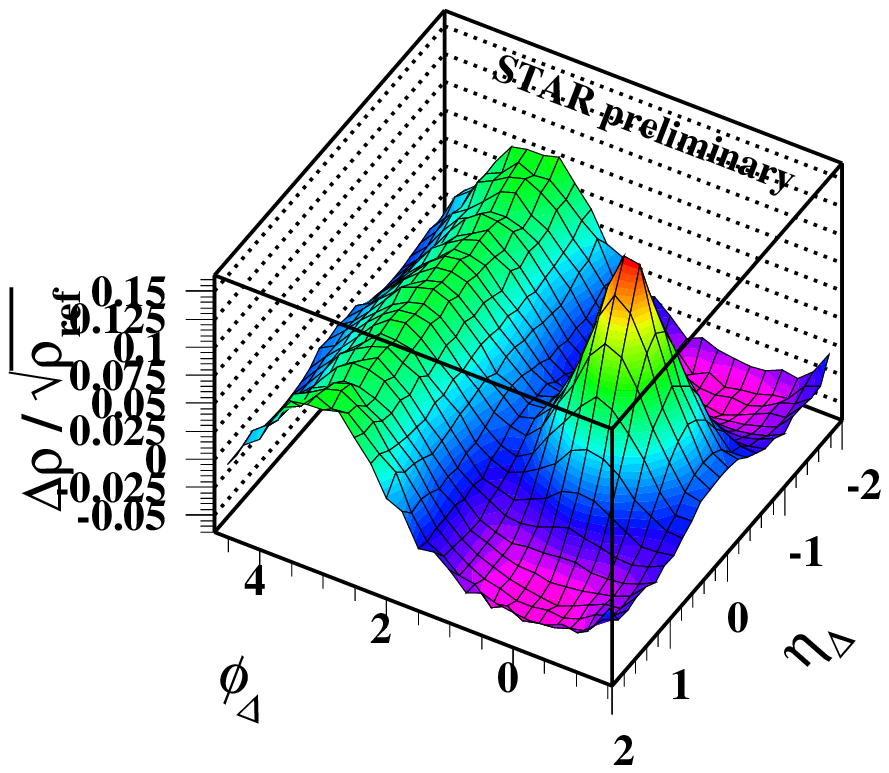} 
\end{center}
\end{minipage}\hspace{0pc}%
\caption{\label{ppref}  Single-particle hard-component distributions on transverse rapidity for ten multiplicity classes of 200 GeV p-p collisions; two-particle distribution on transverse rapidity and soft and hard components respectively of joint autocorrelations on pseudorapidity and azimuth. 
}
\end{figure}

The reference system for low-$Q^2$ partons in A-A collisions is the {\em hard component} of correlations in p-p collisions. The single-particle $p_t$ spectrum for p-p collisions can be decomposed into soft and hard components on the basis of event multiplicity dependence~\cite{jeffismd}. Event multiplicity determines statistically the fraction of p-p collisions containing {\em observable} parton scattering (hard component). Hard components for ten multiplicity classes in the first panel of Fig.~\ref{ppref}, obtained by subtracting fixed soft-component spectrum model $S_0$, are plotted on transverse rapidity $y_t \equiv \ln\{(m_t + p_t)/m_0\}$. The approximately gaussian distributions on $y_t$ may be compared with conventional fragmentation functions plotted on logarithmic variable $\xi \equiv \ln\{E_{jet} / p_t\}$~\cite{opal}. Such single-particle structures motivated a study of two-particle correlations on $(y_{t1},y_{t2})$. An example in Fig.~\ref{ppref} (second panel) reveals structures at smaller and larger $y_t$.

Soft and hard correlation components on $y_t$, interpreted as longitudinal string fragments (smaller $y_t$) and transverse parton fragments (larger $y_t$), produce corresponding structures in joint angular autocorrelations on $(\eta_\Delta,\phi_\Delta)$. In the third panel, string-fragment correlations for unlike-sign pairs are determined by local charge and transverse-momentum conservation (the sharp peak at the origin is conversion electrons). Minimum-bias parton fragments in the fourth panel produce classic jet correlations, with a same-side ($\eta_\Delta < \pi/2$) jet cone at the origin and an away-side ($\eta_\Delta > \pi/2$) ridge corresponding to the broad distribution of parton-pair centers of momentum. Similar-quality parton fragment distributions on $(\eta_\Delta,\phi_\Delta)$ can be obtained for both $p_t$s of a hadron pair down to 0.35 GeV/c (parton $Q/2 \sim$ 1 GeV). 
The criteria for partons as Brownian probes are 1) $Q^2$ large enough that resulting hadron correlations are statistically significant and uniquely assigned to parton fragments, and 2) $Q^2$ small enough that correlations are significantly modified by {\em local} medium dynamics. In the QCD context the medium itself is formed from low-$Q^2$ partons. In the low-$Q^2$ regime `partons' may not interact as {\em point} color charges, and complex couplings to the medium, {\em e.g.,} tensor components of the velocity field (Hubble expansion), may be important. Non-perturbative aspects of low-$Q^2$ parton collisions should be accessible {\em via} low-$p_t$ fragment angular autocorrelations and two-particle $y_t$ distributions.


\section{$\langle p_t \rangle$ fluctuations and prehadronic temperature/velocity structure}

\begin{figure}[h]
\begin{minipage}{38pc} \hfil
 \includegraphics[width=7.5pc,height=8pc]{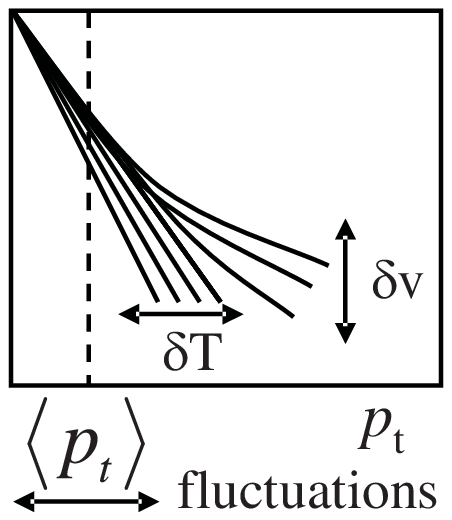} \hfil 
\includegraphics[width=10pc,height=8pc]{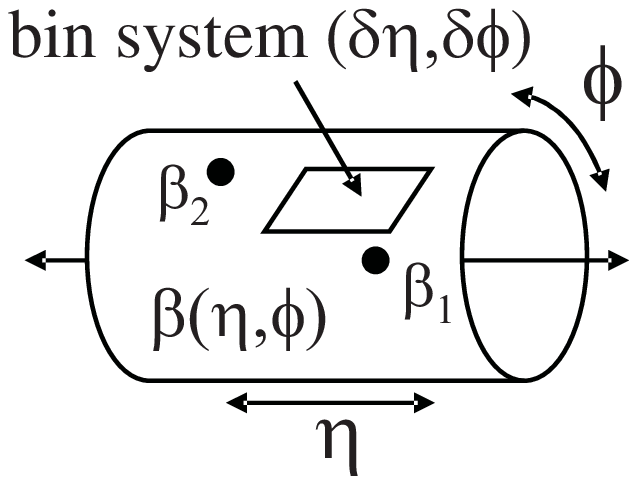} \hfil
\includegraphics[width=7.5pc,height=8pc]{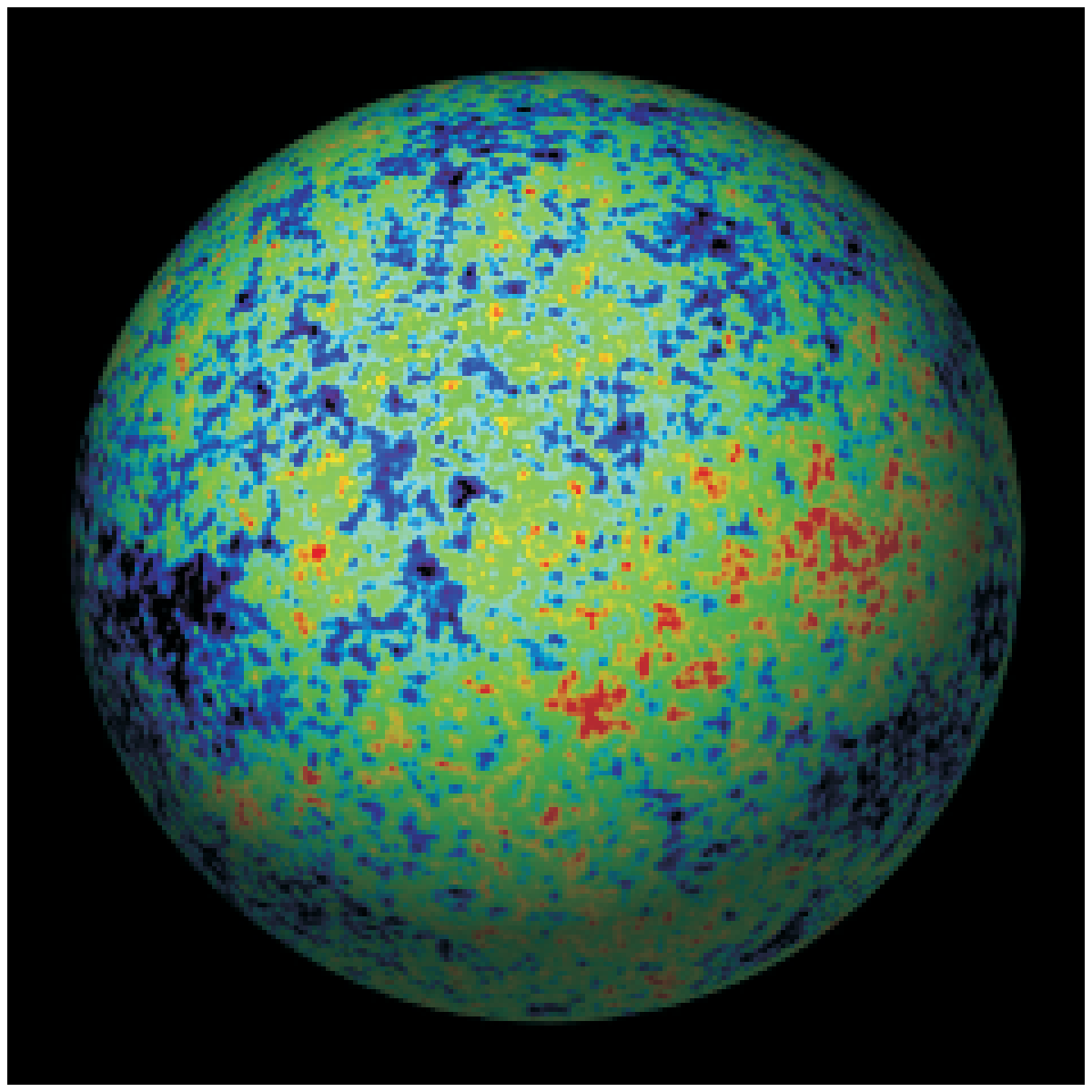}  \hfil
 \includegraphics[width=7.5pc,height=8pc]{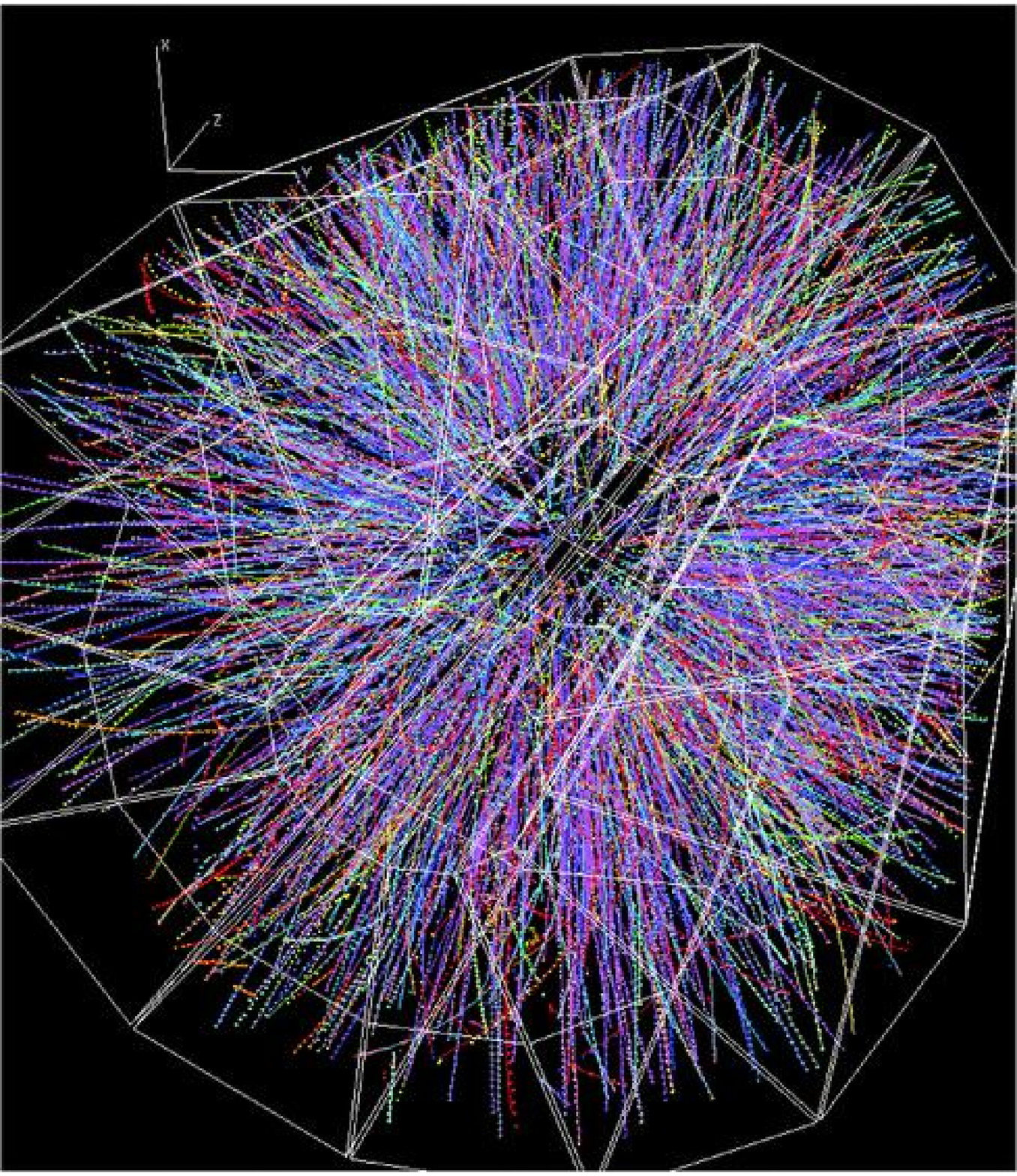}  \hfil
\end{minipage}\hspace{0pc}%
\caption{\label{ptflucts}  Sources of $\langle p_t \rangle$ fluctuations, temperature/velocity ($\beta$) distribution on binned angle space, WMAP microwave distribution and particle distribution from central Au-Au collision.}
\end{figure}

Event-wise $\langle p_t \rangle$ fluctuations generally result from {\em local} event-wise changes in the shape of the single-particle $p_t$ spectrum, as illustrated in Fig.~\ref{ptflucts} (first panel). In each collision, a distribution of `source' temperature and/or velocity on $(\eta,\phi)$ determines the {\em local parent} $p_t$ spectrum shape. Each hadron $p_t$ samples a spectrum shape determined by the sample location, as shown in Fig.~\ref{ptflucts} (second panel). The local parent shape can be characterized schematically by parameter $\beta(\eta,\phi)$, interpreted loosely as $1/T$ or $v/c$ for the local pre-hadronic medium. Variation of either or both parameters relative to an ensemble mean results in $\langle p_t \rangle$ fluctuations.

A similar situation is encountered in studies of the cosmic microwave background (CMB) as shown in Fig.~\ref{ptflucts} (third panel)~\cite{wmap}. The temperature distribution $\beta$  on the unit sphere is represented by the microwave power density (local spectrum integral rather than mean). The $\beta(\theta,\phi)$ structure for that single event is directly observable due to large photon numbers. In contrast, for a single heavy ion collision as in Fig.~\ref{ptflucts} (fourth panel) the parent distribution is sparsely sampled by $\sim 1000$ final-state hadrons, and parent properties are not accessible on an event-wise basis. 

Interpreting $\langle p_t \rangle$ fluctuations has two aspects: 1) study equivalent two-particle number correlations on $p_t$ or $y_t$, which reveal medium modification of the two-particle parton fragment distribution---those correlations are directly related to a distribution on $(\beta_1,\beta_2)$ sensitive to in-medium parton dissipation; 2) invert the {\em scale or bin-size dependence} of $\langle p_t \rangle$ fluctuations to obtain $p_t$ autocorrelations on $(\eta,\phi)$ which reveal details of event-wise $\beta(\eta,\phi)$ distribution. We first consider properties of $\beta(\eta,\phi)$ as a random variable and its relation to two-particle correlations on $p_t$ or $y_t$. We then employ $p_t$ autocorrelations from  $\langle p_t \rangle$ fluctuations to infer aspects of the $\beta(\eta,\phi)$ distribution which depend only on separation of pairs of points on $(\eta,\phi)$. 


\section{Parton Dissipation in the A-A Medium}

$\langle p_t \rangle$ fluctuations can be related to a 1D distribution on temperature/velocity parameter $\beta$ and corresponding {\em two-point} distribution on $(\beta_1,\beta_2)$. Each entry of those distributions corresponds to an event-wise $p_t$ spectrum in a single bin or pair of bins on $(\eta,\phi)$. The frequency distribution on $\beta$ represents variation of the single-particle $p_t$ spectrum shape. For Gaussian-random fluctuations the relative variance of the $\beta$ distribution is $\sigma^2_\beta / \beta_0^2 \equiv 1/n$, where $n$ is the exponent of L\'evy distribution $A / (1 + \beta_0 (m_t - m_0)/n)^n$ describing the average $p_t$ spectrum shape~\cite{levy}. The shape of the single-particle spectrum is thus related to the event-wise temperature/velocity distribution. Other aspects of shape determination, such as collective radial flow, also contribute to exponent $n$. We therefore consider the two-point distribution on $(\beta_1,\beta_2)$. 


Given the correspondence between the fluctuation distribution on $\beta$ and the shape of the single-particle spectrum on $p_t$ we seek the relation between the distribution on $(\beta_1,\beta_2)$ and the shape of the two-particle distribution on $(p_{t1},p_{t2})$. The distribution on $(\beta_1,\beta_2)$ provides information about the correlation structure of event-wise $\beta$ distributions. The two-particle L\'evy distribution on $(p_{t1},p_{t2})$, constructed as a Cartesian product of two single-particle distributions with L\'evy exponent $n$, represents a mixed-pair reference distribution (pairs from different but similar events). We can also define a  two-particle object L\'evy distribution representing sibling pairs (pairs formed from single events), with exponents $n_\Sigma$ and $n_\Delta$ representing variances on sum and difference axes $(\beta_\Sigma,\beta_\Delta)$. The ratio of object and reference distributions reveals a saddle-shaped structure whose curvatures measure temperature/velocity correlations on $(\eta,\phi)$.


\begin{figure}[h]
\begin{minipage}{38pc} \hfil
 \includegraphics[width=8pc,height=8pc]{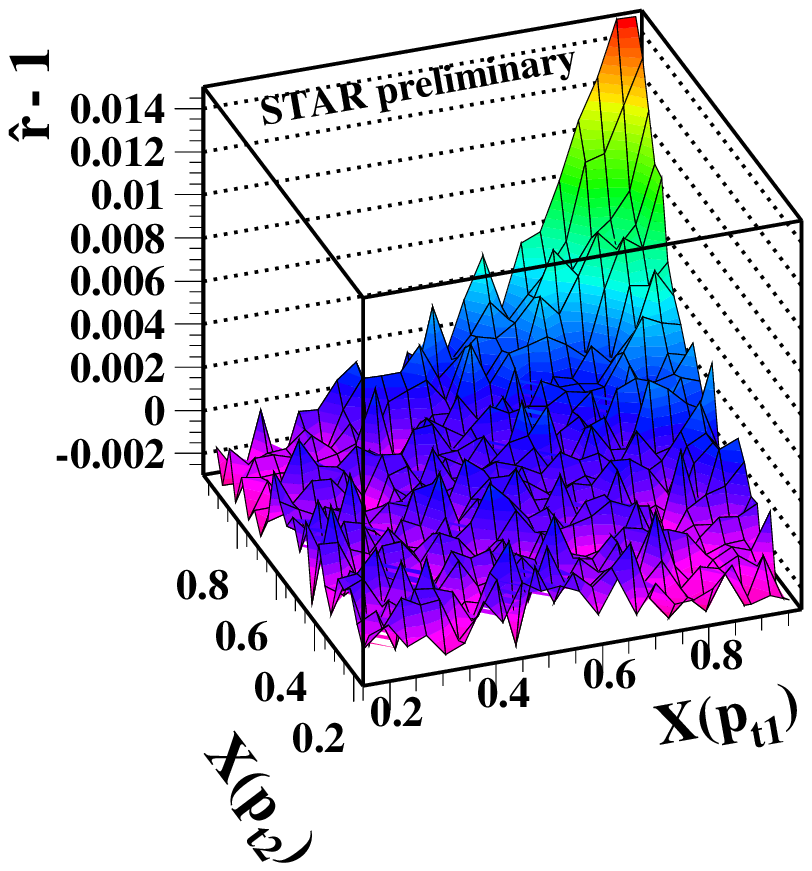} \hfil 
\includegraphics[width=8pc,height=8pc]{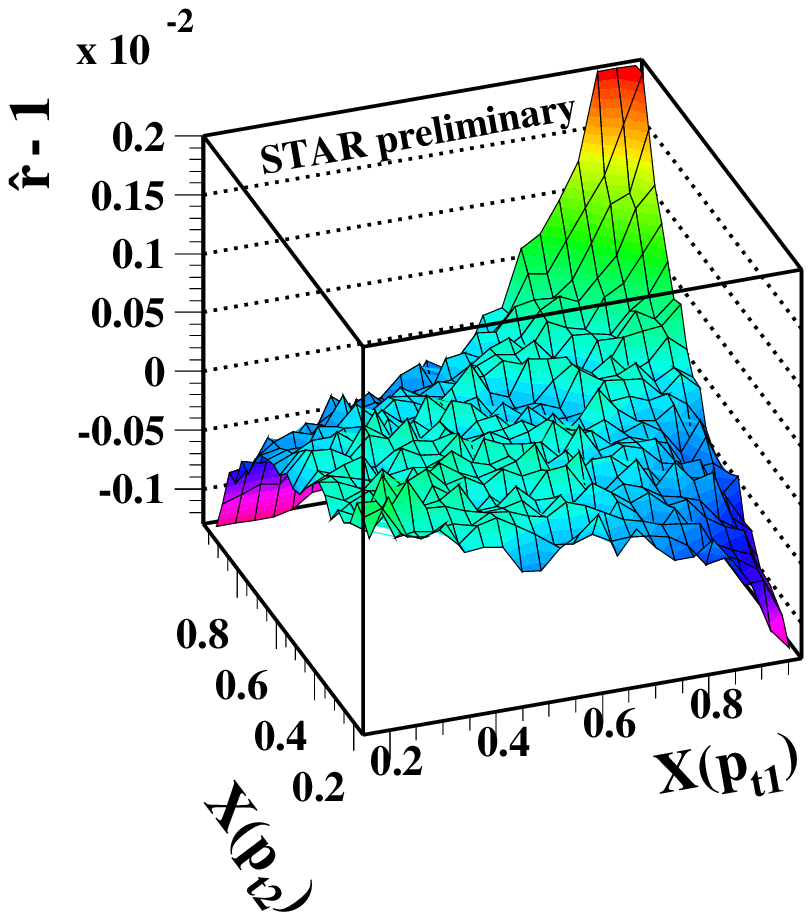} \hfil
 \includegraphics[width=9pc,height=8pc]{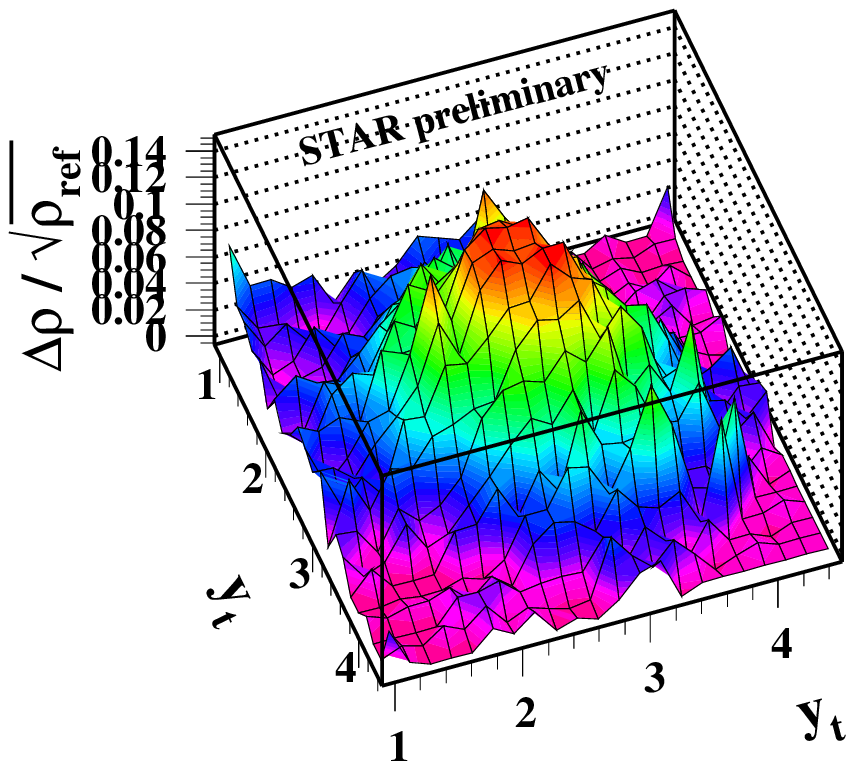} \hfil 
\includegraphics[width=9pc,height=8pc]{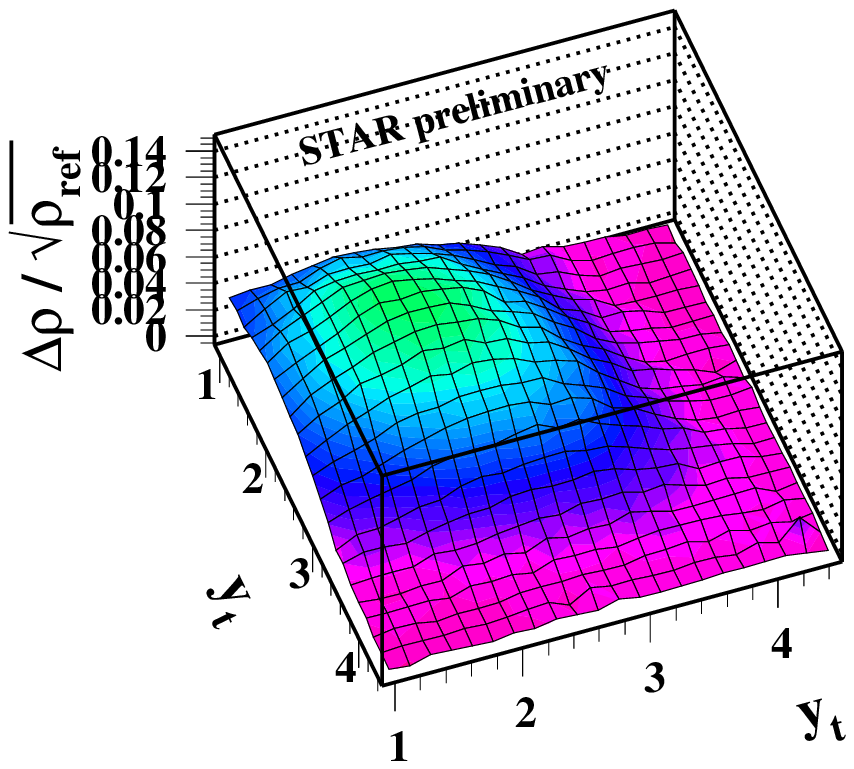} \hfil
\end{minipage}\hspace{0pc}%
\caption{\label{partdiss}  Left panels: Two-particle distributions on $X(p_t)$ (see text) for peripheral and central Au-Au collisions. Right panels: Corresponding number distributions on transverse rapidity.}
\end{figure}  

Ratios of sibling to mixed pair densities for 130 GeV Au-Au collisions are shown in Fig.~\ref{partdiss} (first two panels) plotted on variable $X(p_t)$~\cite{mtxmt}. Those panels are dominated by a {\em L\'evy saddle}, a 2D manifestation of two-particle $p_t$ spectrum shape variation due to velocity and temperature fluctuations in the parent distribution. The saddle is an intermediate shape in the dissipation process; its curvatures reflect the correlation structure of the $(\beta_1,\beta_2)$ distribution, especially its covariance as discussed in~\cite{mtxmt}. The saddle curvatures on sum and difference variables, measured by $1/n_\Sigma - 1/n$ and $1/n_\Delta - 1/n$, represent the variance excesses (beyond independent $p_t$ sampling from a fixed parent) and covariance of temperature/velocity fluctuations for small-amplitude Gaussian-random fluctuations. For an equilibrated system the saddle would be flat (zero curvatures), and $\beta$ fluctuations would be consistent with finite-number fluctuations: $\sigma^2_{\beta_\Sigma} / \beta_0^2 = \sigma^2_{\beta_\Delta} / \beta_0^2 = 1/n$. The integral of correlations on $(p_{t1},p_{t2})$, measured by the saddle-curvature difference $1/n_\Sigma - 1/n_\Delta$, is equivalent to $\langle p_t \rangle$ fluctuations measured in the corresponding detector acceptance~\cite{ptprl}. With increasing Au-Au centrality the curvature on the difference axis increases strongly, while that on the sum axis approaches zero~\cite{mtxmt}.


More recently, we have transitioned from {\em per-pair} correlation measure $\hat r - 1$ plotted on variable $X(p_t)$ to {\em per-particle} density ratio $\Delta \rho / \sqrt{\rho_{ref}}$ plotted on transverse rapidity $y_t$. We wish to follow, within a single context, the transition from parton fragment distributions in elementary collisions to correlations from parton dissipation in a bulk medium. Fig.~\ref{ptflucts} (last two panels) shows $\Delta \rho / \sqrt{\rho_{ref}}$ on $(y_{t1},y_{t2})$ for peripheral and central Au-Au collisions at 200 GeV. The logarithmic $y_t$ interval [1,4.5] corresponds to linear $p_t \sim$ [0.15,6] GeV/c. Peripheral collisions produce a 2D minimum-bias parton fragment distribution peaked at $y_t \sim 2.5$ ($p_t \sim 1$ GeV/c), similar to p-p collisions but without small-$y_t$ correlations from string fragmentation. As centrality increases the fragment distribution is transported to smaller $y_t$ and approaches a shape corresponding to the L\'evy saddle on $X(p_t) \times X(p_t)$. In this format we can study the transition with A-A centrality between two extreme cases: 1) {\em in vacuo} distributions of string and parton fragments and 2) gaussian-random variation of $\beta$ on $(\eta,\phi)$ for a nearly-equilibrated system. Parton dissipation in the A-A bulk medium is represented by the transition between those extremes. 

\section{$\langle p_t \rangle$ fluctuations and $p_t$ autocorrelations}

The previous section describes $\langle p_t \rangle$ fluctuations in terms of two-particle number densities on $(p_{t1},p_{t2})$ or its logarithmic equivalent $(y_{t1},y_{t2})$, the issue being modification of the two-particle parton fragment distribution with changing A-A centrality. One can also express $\langle p_t \rangle$ fluctuations in terms of two-particle $p_t$ distributions on $(\eta,\phi)$ which reveal different aspects of the underlying two-particle number distribution on vector momentum. This section describes a procedure to determine the correlation structure of the $\beta(\eta,\phi)$ distribution as a temperature/velocity distribution on the prehadronic medium.

Fluctuations in bins of a given size or scale are determined by two-particle correlations with characteristic lengths less than or equal to the bin scale. By measuring fluctuation magnitudes as a function of bin size one can recover some details of the two-particle correlation structure---those aspects which depend on the separation of pairs of points, not on their absolute positions. The relation between fluctuations and correlations is given by the integral equation~\cite{inverse}
\bea \label{eq1}
\Delta \sigma^2_{p_t:n}(m \, \epsilon_\eta, n \, \epsilon_\phi) 
= 4  \sum_{k,l=1}^{m,n} \epsilon_\eta \epsilon_\phi & &\hspace{-.25in} K_{mn;kl}  \,   \frac{\Delta \rho(p_t:n;k\,\epsilon_\eta, l\, \epsilon_\phi) }{ \sqrt{\rho_{ref}(n;k\,\epsilon_\eta, l\, \epsilon_\phi)}} ,
\end{eqnarray}
with kernel $K_{mn;kl} \equiv (m - {k + 1/2})/{m} \cdot (n-{l+1/2})/{n}$ representing the 2D macrobin system, $\Delta \sigma^2_{p_t:n}(\delta \eta,\delta \phi)$ is a variance excess and $\Delta \rho(p_t:n;)  / \sqrt{\rho_{ref}(n)}$ is an autocorrelation density ratio. That equation can be inverted numerically to obtain the $p_t$ autocorrelation.

\begin{figure}[h]
\begin{minipage}{38pc}
\includegraphics[width=9pc,height=8pc]{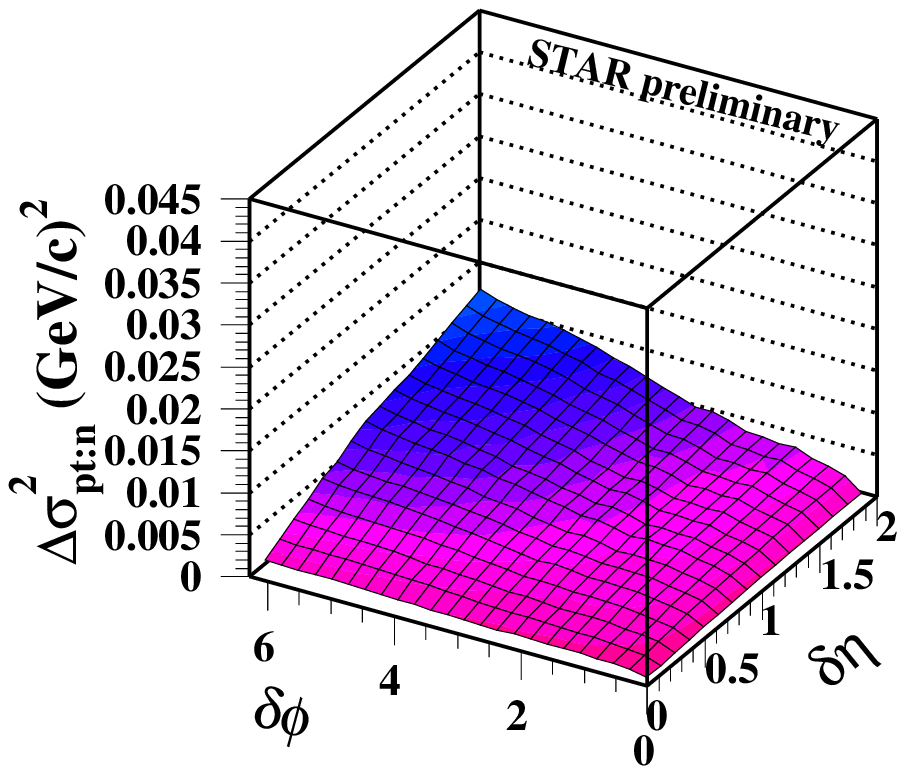}
 \includegraphics[width=9pc,height=8pc]{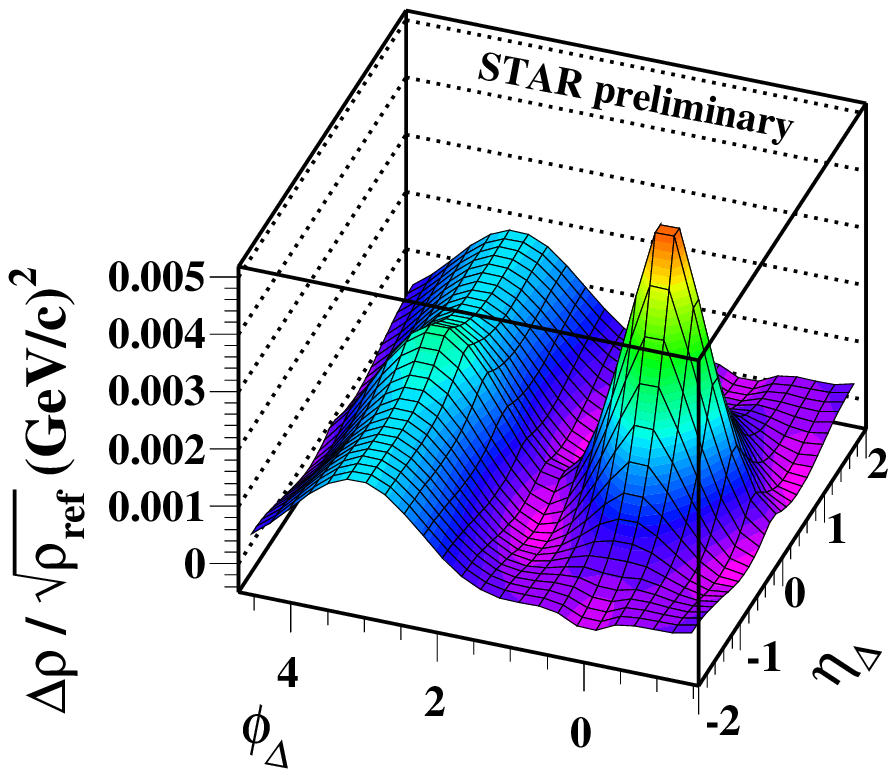} 
\includegraphics[width=9pc,height=8pc]{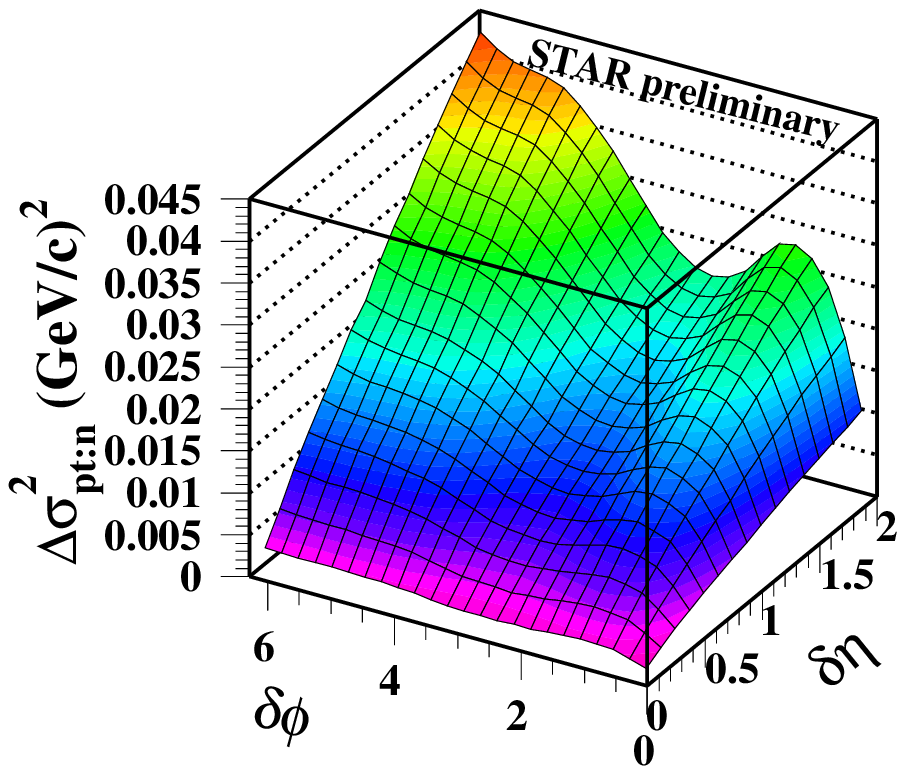}
 \includegraphics[width=9pc,height=8pc]{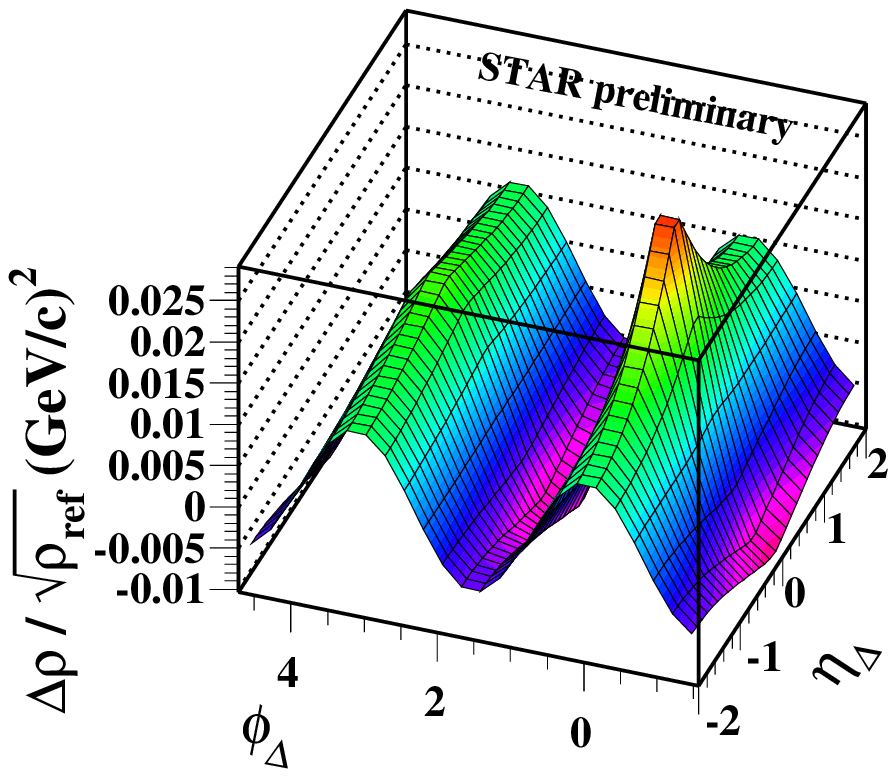} 
\end{minipage}\hspace{0pc}%
\caption{\label{fluctcorr}  Left panels: $\langle p_t \rangle$ fluctuation scale dependence and corresponding $p_t$ autocorrelation for peripheral Au-Au collisions. Right panels: Similar distributions for mid-central collisions.}
\end{figure}  

Fig.~\ref{fluctcorr} shows fluctuation scale dependence on bin sizes $(\delta \eta,\delta \phi)$ and joint $p_t$ autocorrelations on difference variables $(\eta_\Delta,\phi_\Delta)$ for peripheral (left panels) and mid-central (right panels) Au-Au collisions. Fluctuation measurements at the full STAR acceptance correspond to the single points at the apex of the distributions on scale. Measurements with different detectors correspond to different regions of those surfaces. Inversion to autocorrelations provides physical interpretation of fluctuation scale dependence. By inverting $\langle p_t \rangle$ fluctuations, parton fragment distributions are visualized as temperature/velocity structures on $(\eta,\phi)$ complementary to number correlations on $(y_{t1},y_{t2})$ described in the previous section. A more comprehensive picture of parton scattering, dissipation and fragmentation is thereby established.




\begin{figure}[h]
\begin{minipage}{19pc}
\begin{center}
 \includegraphics[width=9pc,height=8pc]{ptautoyes200-0xyz} 
\includegraphics[width=9pc,height=8pc]{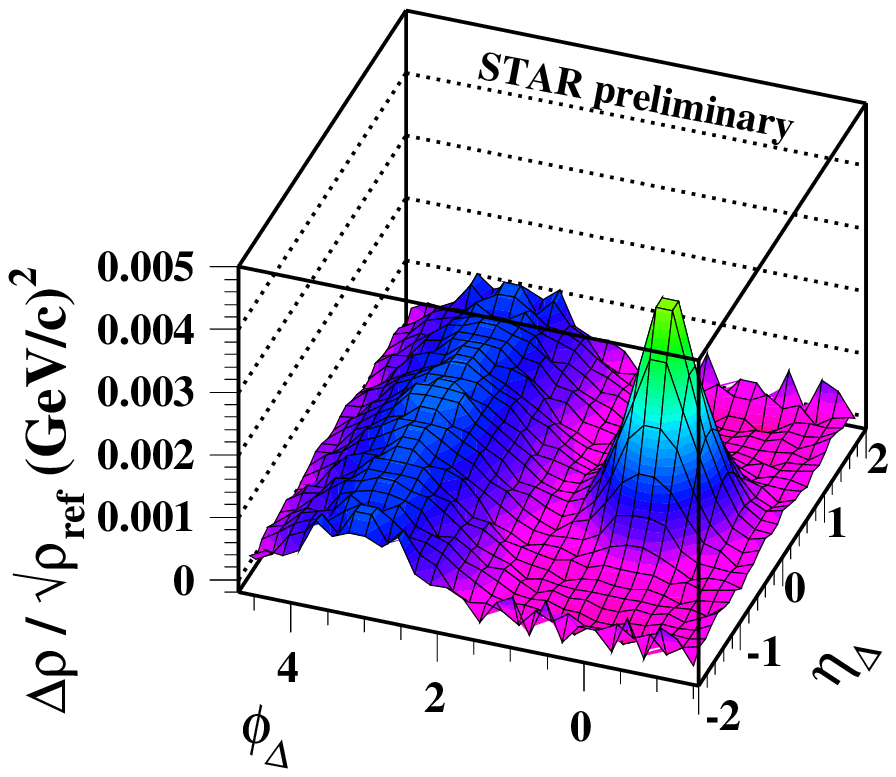}
\end{center} 
\end{minipage}
\hfil
\begin{minipage}{19pc}
\begin{center}
\includegraphics[width=9pc]{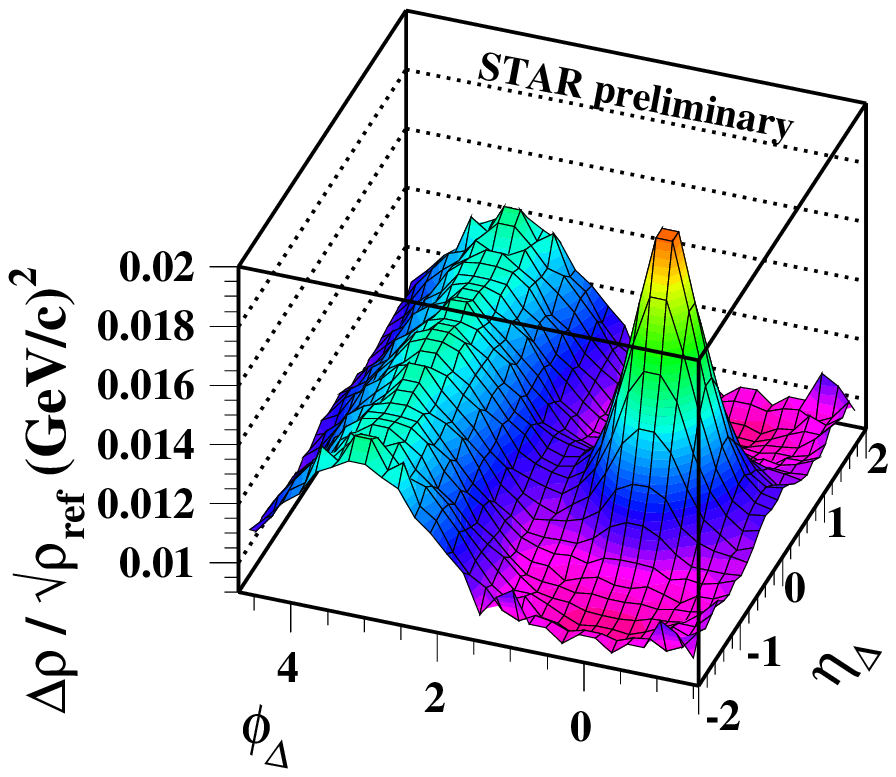} 
\includegraphics[width=9pc]{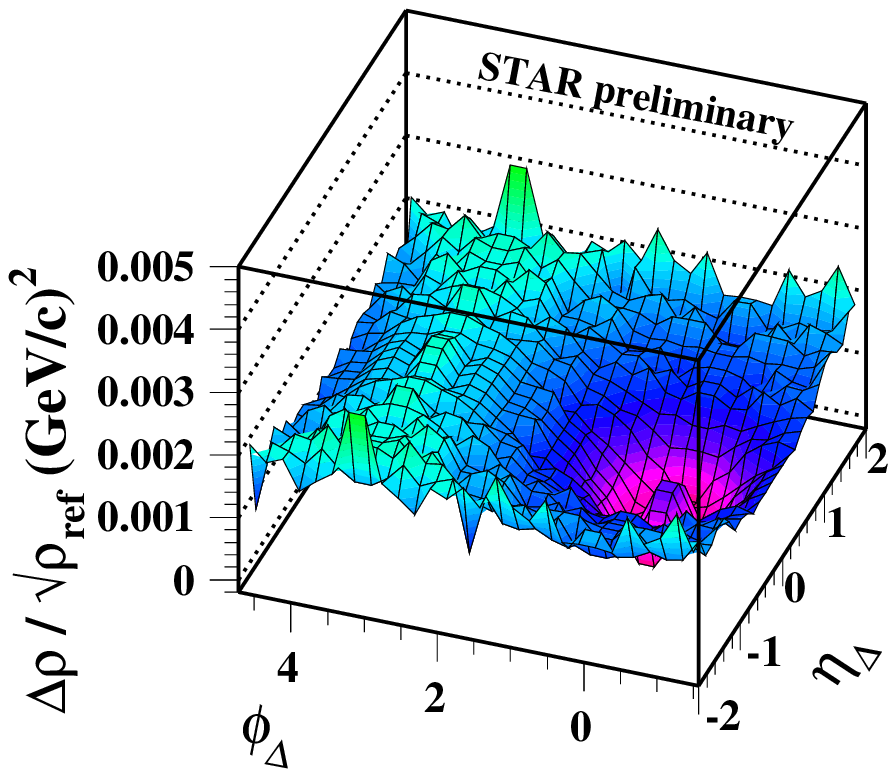} 
\end{center}
\end{minipage}\hspace{0pc}%
\caption{\label{dirauto} $p_t$ autocorrelation from inversion for peripheral Au-Au collisions, autocorrelations from pair counting for minimum-bias p-p collisions and from collisions with $n_{ch} \geq 9$.}
\end{figure}  

$p_t$ autocorrelations can also be determined directly by pair counting. In Fig.~\ref{dirauto} the peripheral Au-Au result from the previous section (first panel) is compared to the minimum-bias p-p result (second panel) and to p-p collisions with $n_{ch} \geq 9$ (third panel). The last panel shows the charge-dependent (like-sign $-$ unlike-sign pairs) $p_t$ autocorrelation for the same event class, reflecting charge-ordering along the jet thrust axis during parton fragmentation. This is the first determination of $p_t$ correlations in p-p collisions.




\section{Local velocity structure and same-side recoil}

\begin{figure}[h]
\begin{minipage}{19pc}
\begin{center}
 \includegraphics[width=19pc,height=8pc]{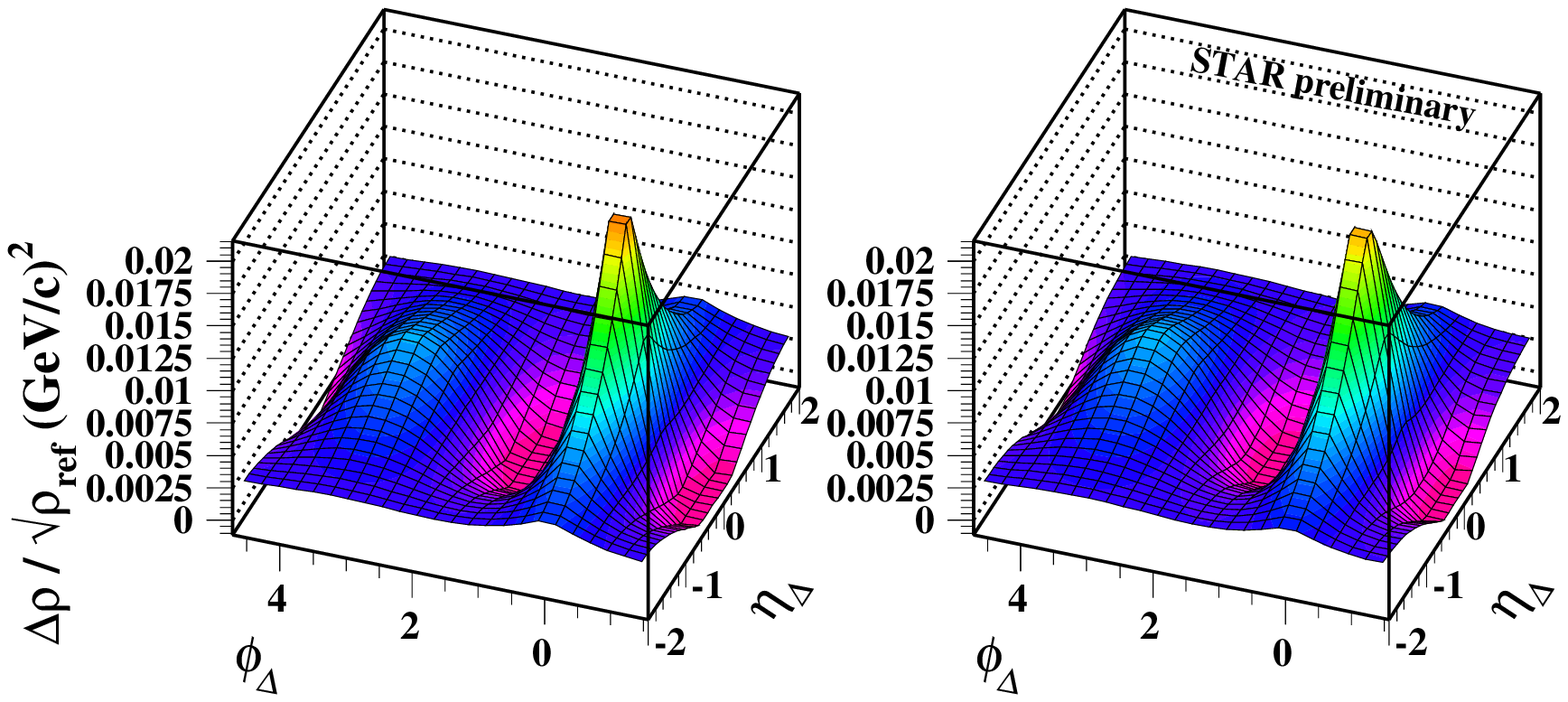} 
\end{center} 
\end{minipage}
\hfil
\begin{minipage}{19pc}
\begin{center}
 \includegraphics[width=10pc,height=8pc]{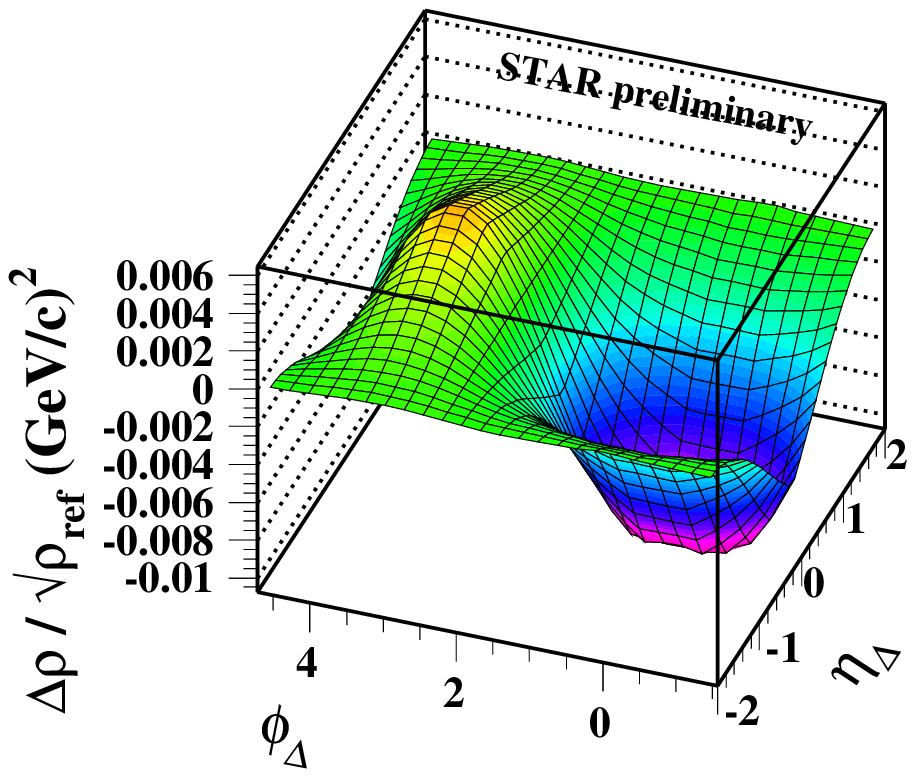} 
\includegraphics[width=7pc,height=7.5pc]{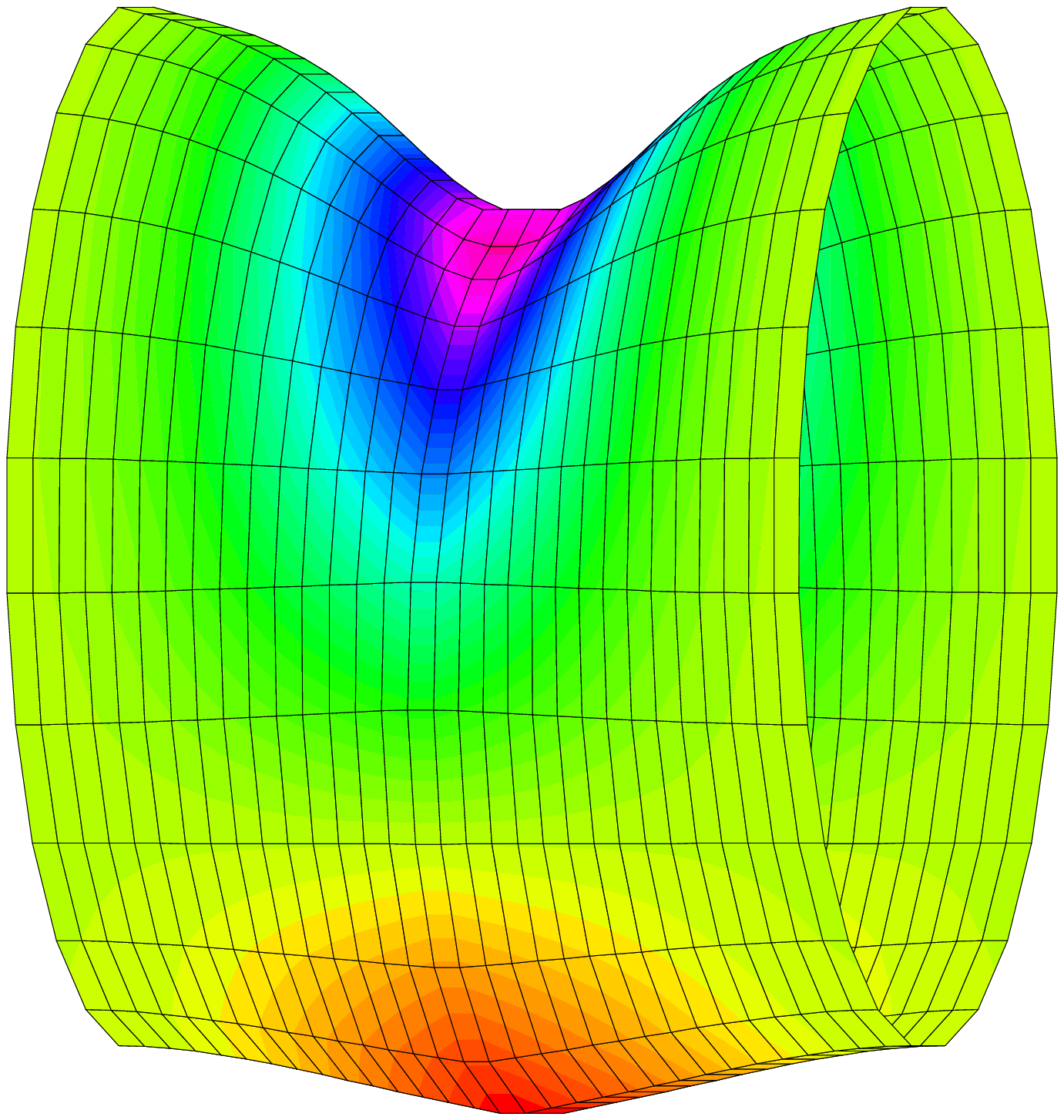} 
\end{center}
\end{minipage}\hspace{0pc}%
\caption{\label{recoil}  $p_t$ autocorrelation for mid-central Au-Au collisions, model fit, data autocorrelation with positive same-side model peak subtracted and the same distribution in cylinder format.}
\end{figure}

Whether derived from pair counting or from fluctuation inversion, the resulting $p_t$ autocorrelations can be separated into several components. We first subtract multipoles on azimuth (azimuth sinusoids independent of pseudorapidity), revealing structure associated with parton scattering and fragmentation. Fig.~\ref{recoil} shows the resulting $p_t$ autocorrelation for 20-30\% central Au-Au collisions at 200 GeV (first panel) and a three-component model fit to that distribution (second panel) including a same-side ($\phi_\Delta < \pi / 2$) positive peak, a same-side negative peak and an away-side ($\phi_\Delta > \pi / 2$) positive peak. The fit is excellent, with residuals at the percent level.
The third panel shows the result of subtracting the positive same-side model peak (representing parton fragments) from the data in the first panel. The shape of the negative same-side peak is very different from the positive peak; there is thus negligible systematic coupling in the fit procedure. The fourth panel shows the data distribution in the third panel plotted in a cylindrical format, suggesting an interpretation in terms of temperature/velocity correlations. 

Histogram values of the $p_t$ autocorrelation effectively measure correlations (covariances) of blue or red shifts of local $p_t$ spectra relative to the ensemble mean spectrum at pairs of points separated by ($\eta_\Delta,\phi_\Delta$). The negative same-side peak can therefore be interpreted as a systematic red shift of local $p_t$ distributions {\em adjacent to} the positive fragment peak. The red shift can in turn be interpreted as recoil of the bulk medium in response to stopping the parton {\em partner} of the observed parton (positive same-side peak). This detailed picture of parton dissipation, stopping and fragmentation in a A-A collisions, including recoil response of the dissipative bulk medium suggested in the fourth panel, is accessed for the first time with joint $p_t$ autocorrelations.


\section{Reconstructing Event-wise Temperature/velocity Structure}

\begin{figure}[h]
\begin{minipage}{18pc}
\begin{center}
 \includegraphics[width=8pc]{wmap-globe} 
\includegraphics[width=9pc,height=8pc]{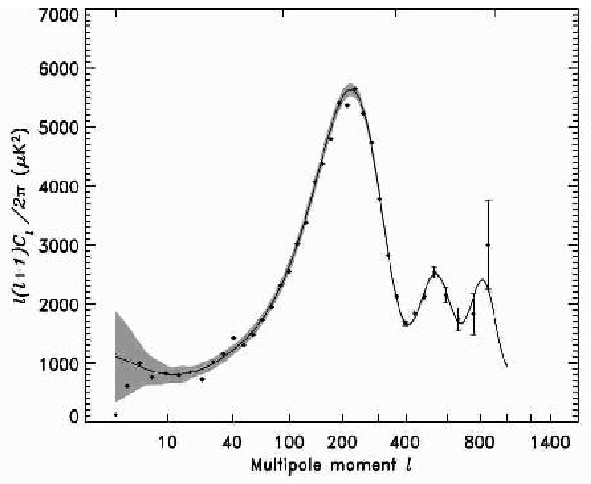}
\end{center} 
\end{minipage}
\hfil
\begin{minipage}{18pc}
\begin{center}
 \includegraphics[width=8.5pc,height=7.5pc]{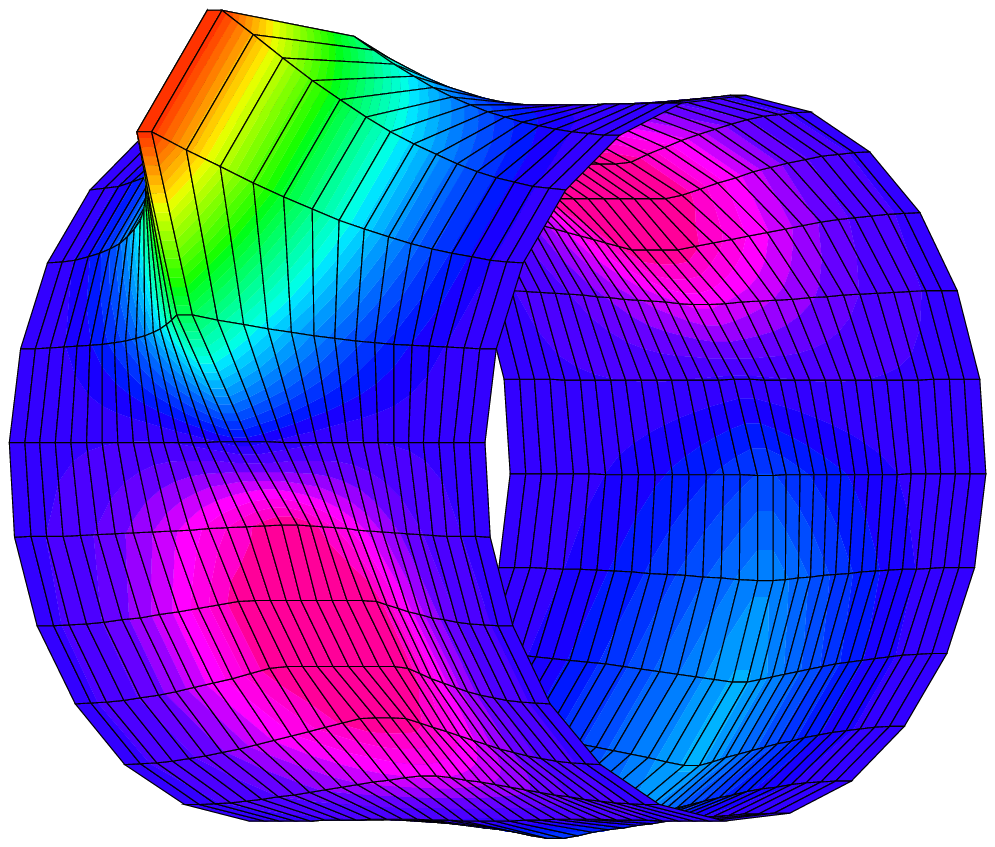} 
\includegraphics[width=8.5pc,height=8pc]{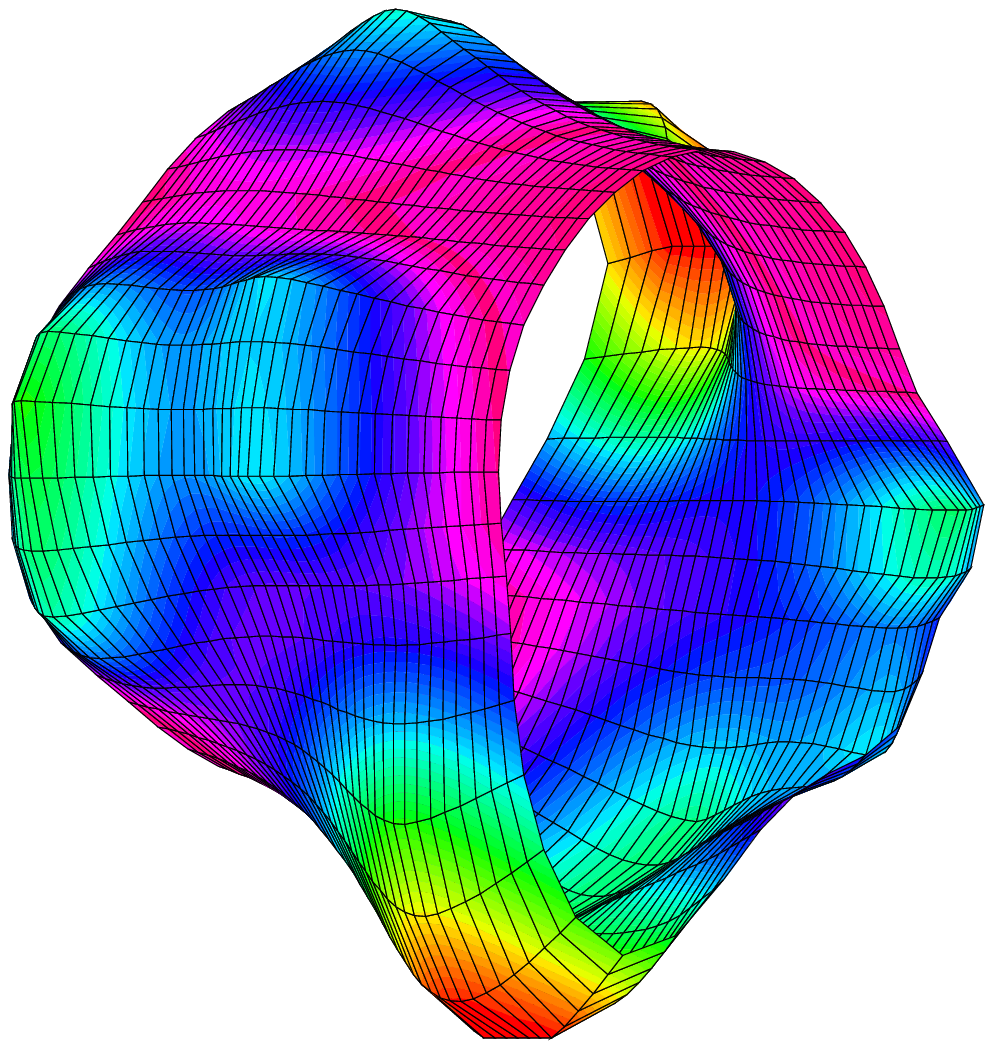} 
\end{center}
\end{minipage}\hspace{0pc}%
\caption{\label{tempvel}  CMB power distribution, corresponding power spectrum, joint $p_t$ autocorrelation for mid-central Au-Au collisions and simulated temperature/velocity distribution for single collision.}
\end{figure}

We now consider the relation of $p_t$ autocorrelations to individual collision events. In Fig.~\ref{tempvel} we repeat the WMAP CMB distribution of microwave power on the unit sphere, picturing a single Big Bang `event' which has a large statistical depth and can therefore be directly observed~\cite{wmap}. Information relevant to cosmological theory is extracted as a power spectrum on polar angle (second panel), formally equivalent (within a Fourier transform) to an autocorrelation according to the Wiener-Khinchine theorem. In some studies, CMB angular autocorrelations and cross-correlation have been determined directly~\cite{cobe}.

In our study of heavy ion collisions we obtain angular $p_t$ autocorrelations as in the third panel. Due to sparse sampling we cannot directly visualize the temperature/velocity structure of individual collision events as for the CMB survey. The local microwave power density of the CMB survey is analogous to local $\langle p_t \rangle$ in a Au-Au collision. For individual collisions, and especially for smaller bin sizes, the event-wise mean values are not significant. However, given  $p_t$ autocorrelations we can simulate event-wise velocity/temperature distributions. We estimate the number of hard parton scatters within the STAR acceptance in a central Au-Au collision as 20-40, based on an analysis of p-p collisions~\cite{jeffismd}. Combining that frequency estimate with shape information from the autocorrelation, and introducing some statistical variation of peak structure about the autocorrelation mean value, we can produce simulated events as shown in Fig.~\ref{tempvel} (fourth panel): distributions on primary angle variables $(\eta,\phi)$, whereas the autocorrelation is on difference variables $(\eta_\Delta,\phi_\Delta)$. This exercise illustrates that while Au-Au collisions are RHIC may be {\em locally equilibrated} prior to kinetic decoupling, they remain {\em highly structured} due to copious parton scattering which is not fully dissipated. Access to that structure requires $p_t$ rather than angular or number autocorrelations on $(\eta,\phi)$ to provide the full picture.

\section{Energy dependence of $\langle p_t \rangle$ fluctuations and parton scattering}

\begin{figure}[h]
\begin{minipage}{19pc}
\begin{center}
 \includegraphics[width=8.5pc]{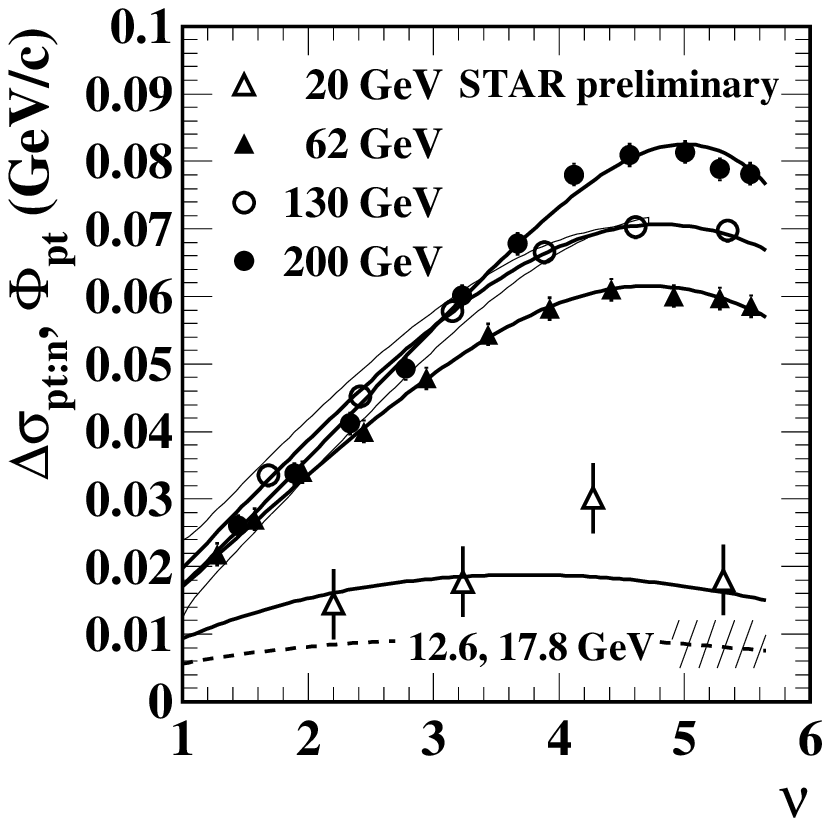} 
\includegraphics[width=8.5pc]{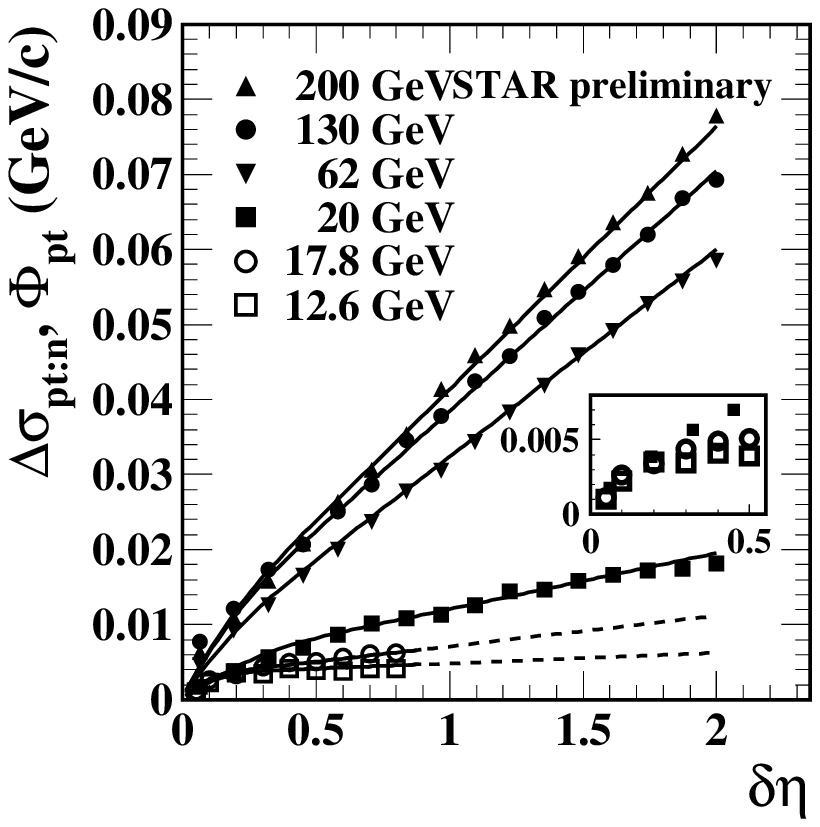}
\end{center} 
\end{minipage}
\hfil
\begin{minipage}{19pc}
\begin{center}
 \includegraphics[width=8.5pc]{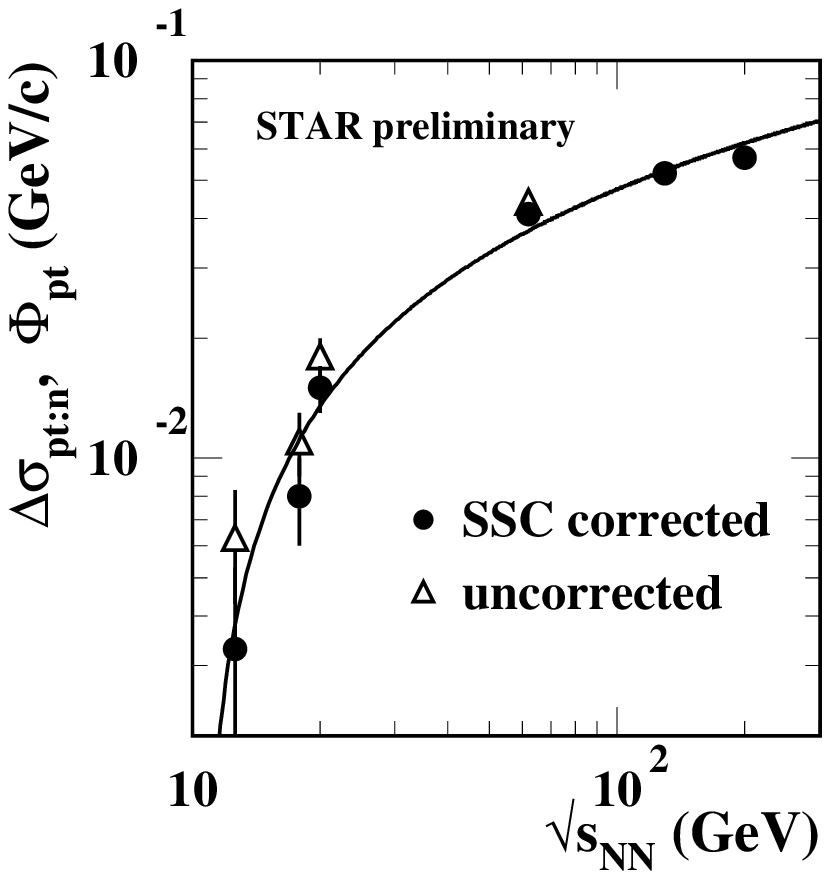} 
\includegraphics[width=8.5pc]{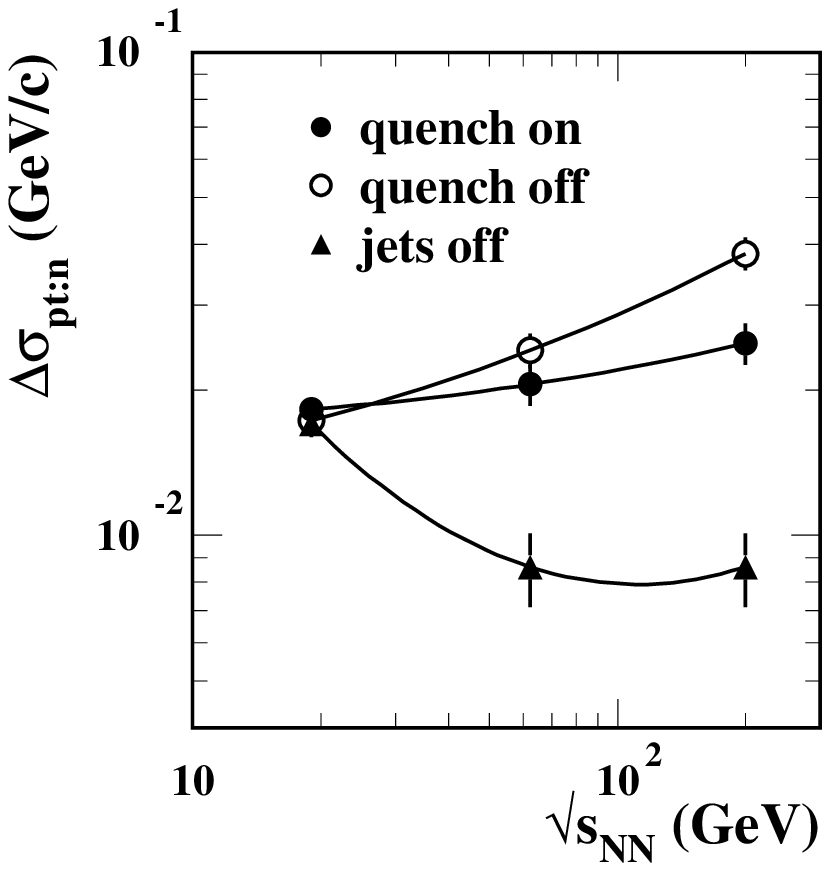} 
\end{center}
\end{minipage}\hspace{0pc}%
\caption{\label{edep}  Centrality, pseudorapidity scale and energy dependence of $\langle p_t \rangle$ fluctuations for central collisions in the STAR acceptance and energy dependence for Hijing Monte Carlo.}
\end{figure}

Given this close connection between parton scattering and fluctuations, the collision-energy dependence of $\langle p_t \rangle$ fluctuations may reveal previously inaccessible parton dynamics at lower collision energies. 
In Fig.~\ref{edep} (first panel) we show the centrality dependence ($\nu$ measures mean participant path length in nucleon diameters) of $\langle p_t \rangle$ fluctuations for four RHIC energies
and a summary (crosshatched region) of SPS fluctuation measurements at 12.6 and 17.3 GeV~\cite{ceres}, all at the full STAR acceptance (CERES measurements are extrapolated). In the second panel the pseudorapidity scale dependence of fluctuations at full azimuth acceptance is shown for central collisions at six energies. Extrapolation of CERES data in the first panel is illustrated by the dashed lines at the bottom of the second. Fluctuation measure $\Delta \sigma_{p_t:n}$ is related to the variance difference in Eq.~(\ref{eq1}) by  $\Delta \sigma^2_{p_t:n} \equiv 2 \sigma_{\hat p_t} \, \Delta \sigma_{p_t:n}$, with $\sigma_{\hat p_t}$ the single-particle variance. To good approximation $\Delta \sigma_{p_t:n} \simeq \Phi_{p_t}$, and both are {\em per particle} fluctuation measures. $\Phi_{p_t}$ was used for the CERES fluctuation measurements.

For either measure we observe a dramatic increase in $\langle p_t \rangle$ fluctuations from SPS to RHIC energies. The centrality dependence in the first panel suggests that fluctuations for p-p and peripheral A-A collisions saturate near 60 GeV, whereas there is monotonic increase for the more central collisions. The scale dependence in the second panel illustrates how measurements with different detector acceptances are related. Measurements over common scale intervals should correspond. At RHIC energies we have demonstrated that $\langle p_t \rangle$ fluctuations are dominated by fragments from low-$Q^2$ parton collisions. The energy dependence of $\Delta \sigma_{p_t:n}$ or $\Phi_{p_t}$ is shown in the third panel of Fig.~\ref{edep}, plotted {\em vs} $\sqrt{s_{NN}}$. We observe that $\langle p_t \rangle$ fluctuations vary almost linearly with $\log\{\sqrt{s_{NN}}/10.5\}$ (solid curve in that panel), suggesting a threshold for {\em observable} parton scattering and fragmentation near 10 GeV. 

Fluctuation measurements based on $\Sigma_{p_t} \simeq \sqrt{\Delta \sigma^2_{p_t:n} / (\bar n_{ch}\, \hat p^2_t)}$~\cite{ceres} appear to contradict the results described here, implying instead negligible energy dependence of $\langle p_t \rangle$ fluctuations from SPS to RHIC. We observe that nuclear collisions at RHIC are dominated by {\em local} temperature/velocity structure from hard parton scattering. $\Sigma_{p_t}$ is a {\em per pair} measure which averages the {\em local} $p_t$ correlation structure dominating RHIC collisions over the {\em entire} detector acceptance, resulting in {\em apparent} reduction of correlations with increasing A-A centrality as $1/N_{participant}$ (per the central limit theorem) and consequent insensitivity to contributions from hard scattering. We want to study {\em separately} the changes in $p_t$ production ($T$) and in the {\em correlation structure} of that produced $p_t$ ($\delta T$) {\em prior to hadronization}. $\Sigma_{p_t}$ by construction estimates relative temperature fluctuations of the form $\delta T / T$. 
It thus divides the structure problem by the production problem, greatly decreasing sensitivity to each. 




\section{Summary}

We have demonstrated that low-$Q^2$ partons, accessed here for the first time by novel analysis techniques including joint autocorrelations, serve as Brownian probes of A-A collisions, being the softest {\em detectable} dynamical objects which experience QCD interactions as color charges. Our analysis of p-p correlations provides an essential reference for A-A collisions. Inversion of the scale dependence of $\langle p_t \rangle$ fluctuations provides the first access to $p_t$ autocorrelations which reveal a complex parton dissipation process in A-A collisions relative to p-p collisions. We observe possible evidence for bulk-medium recoil in response to parton stopping. We also observe strong energy dependence of $\langle p_t \rangle$ fluctuations, which is to be expected given the dominant role of scattered partons in driving those fluctuations.

This work was supported in part by the Office of Science of the U.S. DoE under grant DE-FG03-97ER41020.


\smallskip

\end{document}